\PassOptionsToPackage{square,comma,numbers,sort&compress,super}{natbib}
\documentclass[12pt,preprint]{aastex}
\usepackage{natbib}
\usepackage{gensymb}
\usepackage{float}
\usepackage{graphicx}
\usepackage{color}
\usepackage{amsmath}
\begin{document}
%\setcitestyle{super,comma,sort&compress,square} 

\title{A Study of the Gamma-Ray Burst Fundamental Plane}

\author{Dainotti, M. G. \altaffilmark{1,2,3,4}, Hernandez, X. \altaffilmark{5}, Postnikov, S. \altaffilmark{6}, Nagataki, S. \altaffilmark{7}, O'brien, P \altaffilmark{8}, Willingale, R. \altaffilmark{8}, S. Striegel \altaffilmark{9}}

\altaffiltext{1}{INAF-Istituto di Astrofisica Spaziale e Fisica cosmica, c/o CNR - Area della Ricerca di Bologna, Via Gobetti 101, I-40129 -
  Bologna, Italy; dainotti@iasfbo.inaf.it}
	\altaffiltext{2}{SLAC NATIONAL ACCELERATOR LABORATORY 2575 Sand Hill Road, Menlo Park, CA 94025, USA; dainotti@mailbox.slac.stanford.edu}
	\altaffiltext{3}{Department of Physics \& Astronomy, Stanford University, Via Pueblo Mall 382, Stanford, CA 94305-4060, USA;
  mdainott@stanford.edu}
	\altaffiltext{4}{Obserwatorium Astronomiczne, Uniwersytet Jagiello\'nski, ul. Orla 171, 31-501 Krak{\'o}w, Poland; dainotti@oa.uj.edu.pl}
\altaffiltext{5}{Instituto de Astronom\'{\i}a, Universidad Nacional Aut\'onoma de M\'exico, Ciudad de M\'exico 04510, M\'exico E-mail: xavier@astro.unam.mx}
\altaffiltext{6}{The Center for Exploration of Energy and Matter, Indiana University, Bloomington, IN 47405, USA; postsergey@gmail.com}
\altaffiltext{7}{RIKEN, Hirosawa, Wako Saitama, Japan; shigehiro.nagataki@riken.jp}
\altaffiltext{8}{Department of Physics \& Astronomy, University of Leicester, Road Leicester LE1 7RH, UK; pto2@leicester.ac.uk; zrw@le.ac.uk}
\altaffiltext{9}{Department of Physics \& Astronomy, San Jose State University, One Washington Square, San Jose, CA 95192, USA; stephanie.striegel@sjsu.edu}

\date{\today}

\begin{abstract}

Long gamma-ray bursts (GRBs) with a plateau phase in their X-ray afterglows obey a 3D relation (Dainotti et al. 2016a), between the rest-frame time at the end of the plateau, $T_a$, its corresponding X-ray luminosity, $L_{a}$, and the peak luminosity in the prompt emission, $L_{peak}$. This 3D relation identifies a GRB fundamental plane whose existence we here confirm.  Here we include the most recent GRBs observed by {\it Swift} to define a `gold sample' (45 GRBs) and obtain an intrinsic scatter about the plane compatible within 1 $\sigma$ with the previous result.
We compare GRB categories, such as short GRBs with extended emission (SEE), X-ray Flashes (XRFs), GRBs associated with SNe (GRB-SNe), a sample of only long-duration GRBs (132), selected from the total sample by excluding GRBs of the previous categories, and the gold sample, composed by GRBs with light curves with good data coverage and relatively flat plateaus. We find that the relation planes for each of these categories are not statistically different from the gold fundamental plane, with the exception of the SSE, which are hence identified as a physically distinct class. The gold fundamental plane has an intrinsic scatter smaller than any plane derived from the other sample categories. Thus, the distance of any particular GRB category from this plane becomes a key parameter. We computed the several category planes with $T_a$ as a dependent parameter obtaining for each category smaller intrinsic scatters (reaching a reduction of $24\%$ for the long GRBs).
The fundamental plane is independent from several prompt and afterglow parameters.

\end{abstract}

\keywords{gamma-rays bursts: general - methods: statistical}
 
\maketitle
\section{Introduction}
Gamma-ray bursts (GRBs) have typical isotropic prompt emission energies, $E_{iso}$, in the range of $10^{53}$ erg, and thus can be
observed up to redshifts, $z$, of $\sim 10$ (Cucchiara et al. 2011). This last feature raises the tantalizing possibility of extending
direct cosmological studies far beyond the redshift range covered by supernovae (SNe). However, GRBs are not simple standard candles,
as their intrinsic energies span several orders of magnitude. The variety of their features makes it extremely difficult to categorize
them under certain common patterns. Indeed, the number of sub-classes into which GRBs are grouped has grown since their discovery.
GRBs are traditionally classified depending on their duration into short ($T_{90} \leq 2$ s) and long ($T_{90} \ge 2$ s) \footnote{
  $T_{90}$ is the time interval over which between $5\%$ and $95\%$ of the total prompt energy is emitted.} \citep{Mazets,Kouveliotou}. Later,
a class of GRBs with mixed properties, such as short GRBs with extended emission (SEE), was discovered \citep{nb06}. Long GRBs,
depending on their fluence (erg cm$^{-2}$), can be divided into normal GRBs or X-ray flashes (XRFs); the latter are empirically
defined as GRBs with a greater fluence in the X-ray band ($2-30$ keV) than in the $\gamma$-ray band ($30 - 400$ keV). In addition,
several GRBs also present associated SNe; hereafter they are referred to as GRB-SNe. Recently, a new category of ultra-long GRBs
has been discovered (Stratta et al. 2013, Nakauchi et al. 2013, Levan et al. 2014). These GRBs present remarkably unusual X-ray and optical light curves, very different from
classical GRBs, with long-lasting highly variable X-ray emission, and optical light curves showing a weak correlation with the behavior
seen in the X-ray. Levan et al. (2014) proposed that these bursts, difficult to detect, are the first examples of a new population
of ultra-long GRBs, which may be astrophysically relatively common. The long durations may be explained by the engine-driven explosions
of stars of much larger radii than those that are usually considered as GRB progenitors, which are thought to have compact Wolf-Rayet
progenitor stars. However, Levan et al. (2014) claimed that it is not possible to unequivocally identify SN signatures within
their light curves or spectra. Thus, they also considered that they may arise from the tidal disruption of stars by supermassive black
holes. Other ultra-long GRBs have been observed, for example, GRB 130925A, to have features that are associated with a low-metallicity blue supergiant progenitor and could characterize the class of ultra-long GRBs \citep{Piro2014}.
%glow spectrum at various epochs. It requires
%an ambient medium with a very low-density wind profile, consistent with that expected from a low-metallicity
%blue supergiant (BSG)

Regarding light-curve morphology, a more complex trend in the afterglow has been observed with the {\it Swift} Satellite
\citep{Gehrels04,OB06} compared to previous missions. Due to {\it Swift}, it has been discovered that there is a flat part, the plateau,
of GRB light curves soon after the steep decay of the prompt emission. Along with these categories, several physical mechanisms for
producing GRBs have also been proposed. For example, the plateau emission has mainly been ascribed to millisecond newborn spinning
neutron stars (e.g., Zhang \& M\'{e}sz\'{a}ros 2001; Toma et al. 2007, Troja et al. 2007; Dall'Osso et al. 2011; Rowlinson et al. 2013,2014;
Rea et al. 2015) or to accretion onto a black hole (Kumar 2008, Cannizzo \& Gerhels 2009; Cannizzo et al. 2011). A very promising field has been
the hunt for correlations between physically meaningful GRB parameters (e.g., Amati et al. 2002; Yonetoku et al. 2004; Ghirlanda et al.
2004; Ghisellini et al. 2009; Oates et al. 2012; Qi et al. 2009; Willingale et al. 2010, , Xu \& Huang 2012, Grupe et al. 2013), in order to employ GRBs as cosmological
indicators, as cosmological tools, and as theoretical model discriminators.

The relations discovered so far suffer from having large scatters (Collazzi \& Schaefer 2008), beyond observational uncertainties, highlighting that the events studied probably come from different classes of systems or perhaps from the same class of objects, but we do not yet observe a sufficiently large number of parameters to characterize the scatter. Indeed, other possible sources of scatter about relations could depend on the difference in bulk Lorenz factor, in the density of the medium that can be homogeneous or of a stellar wind type, in the viewing angles, etc. In addition, the majority of such relations consider the GRB emission as isotropic, but a small jet opening is also possible. The jet opening angle is very difficult to infer owing due to the paucity of multi-wavelength observations. Several methods have been proposed in the literature to obtain an independent estimate of the jet opening angle (Ghirlanda et al. 2004; Burrows et al. 2006, Soderberg et al. 2006, Grupe et al. 2006, Grupe et al. 2007; Lu et al. 2012; Guidorzi et al. 2014, Fong et al. 2015; Goldstein et al. 2016, Troja et al. 2016b).

As was pointed out in Dainotti et al. (2010), in order to properly use GRB relations as reliable model discriminators and cosmological tools, it is necessary to define type-specific GRB categories to yield a more homogeneous, observationally motivated sample. In fact, it may be that the scatter of the GRB scaling relations could be partly explained by the mixing of GRBs with different intrinsic physics. Isolating GRB categories allows us to derive tighter relations, thus increasing the accuracy with which cosmological parameters are inferred (e.g., Cardone et al. 2009, 2010; Dainotti et al. 2013b; Postnikov et al. 2014).

One of the first attempts to standardize GRBs in the afterglow parameters was presented with the Dainotti relation (Dainotti et al. 2008, 2010, 2011a, 2015a, 2017a), where an approximately inversely proportional law between the rest-frame time at the end of the plateau phase, $T_{a}$ (in previous papers $T^{*}_a$), and its corresponding luminosity, $L_{a}$, was discovered. Dainotti et al. (2013a) proved through the robust statistical Efron \& Petrosian (1992) method, hereafter EP, that this correlation is intrinsic, and not an
artifact of selection effects or due to instrumental threshold truncation, as is also the case for the $L_{peak}-L_{a}$ relation (Dainotti et al. 2011b, 2015b), where $L_{peak}$ is the 1 $s$ peak luminosity in the prompt emission. For a review on GRB relations see the following reviews: Dainotti et al. (2016b) and Dainotti \& Del Vecchio (2017).

In this paper we use a large GRB sample to confirm results discussed in Dainotti et al. (2016a), namely, that the peculiar plateau phase in GRBs can be employed
to isolate a sub-class of events that define a very tight plane in a 3D space of $(\log L_{a}, \log T_{a}, \log L_{peak})$. Results of this paper have been presented in a NASA press release\footnote{$https://swift.gsfc.nasa.gov/news/2016/grbs\texttt{\_}std\texttt{\_}candles.html$} at the 228th AAS Meeting.
We here confirm that the scatter about the fundamental plane is the smallest for the gold sample, a specific class of GRBs without
steep plateaus and with good data coverage when we consider $L_a$ as the independent variable. We have extended the previous gold sample to 45 GRBs obtaining an
intrinsic scatter compatible with our previous finding to within 1 $\sigma$. We find that other relation planes for the different
categories have larger scatter than the fundamental plane derived from the gold sample. We also show that the fundamental plane is independent from several prompt and afterglow
parameters, such as the rest-frame prompt emission duration, $\frac{T_{90}}{(1+z)}= T^{*}_{90}$, rest-frame peak energy, $E_{peak}*(1+z)=E^{*}_{peak}$, the temporal decay index after the plateau emission, $\alpha$, and the jet opening angle, $\theta_{jet}$. Thus, we can conclude that the plane is stable and not in fact a hypersurface
  in four dimensions. This analysis is relevant, because it shows the robustness of the fundamental plane, and hence we can possibly use it in the future as a cosmological tool owing to its small scatter. We note that we have already addressed for both the $L_X-T^{*}_a$ and the $L_{peak}-T^{*}_a$ relations their cosmological evolutions and determined their intrinsic correlations.
  This paper also constitutes an update and presents a new investigation with respect to the previous analysis presented in Dainotti et al. (2016a), as the validity of our conclusion has been extended by including a high-energy subsample of GRBs observed by the Fermi Gamma-ray Burst Monitor (GBM). We also test a plane for which the variable $T_a$ is a function of $L_{peak}$ and $L_a$. The choice of $T_a$ as a dependent variable reduces the intrinsic scatter of all the long GRBs by a further $24\%$. In addition, we show that the distance to the gold sample fundamental plane can be considered as a new key parameter to discriminate among long and SEE bursts. In Sections \S \ref{sample selection} and \S \ref{3D correlation} we describe the {\it Swift} data samples used and the three-parameter relations for those samples, respectively. In section \S \ref{results} we present the independence of the $(L_{a}, T_{a}, L_{peak})$ relation from other relevant prompt and afterglow parameters. In \S \ref{discussion}, we summarize our findings and conclusions.

\section{Sample Selection}\label{sample selection}

We analyzed 183 GRB X-ray plateau afterglows detected by {\it Swift} from 2005 January up to 2016 July with known redshifts,
spectroscopic or photometric, available in Xiao \& Schaefer (2009), on the Greiner
web page, \footnote{http://www.mpe.mpg.de/jcg/grbgen.html} and in the Gamma-ray Coordinates Network (GCN) circulars and
notices, \footnote{http://gcn.gsfc.nasa.gov/}, excluding redshifts for which there is only a lower or an upper limit. The redshift
range of our sample is $(0.033, 9.4)$. We include all GRBs for which the Burst Alert Telescope (BAT) + X-Ray Telescope (XRT)
light curves can be fitted by the phenomenological Willingale et al. (2007, hereafter W07) model.
The W07 functional form for $f(t)$ is

\begin{equation}
f(t) = \left \{
\begin{array}{ll}
\displaystyle{F_i \exp{\left ( \alpha_i \left( 1 - \frac{t}{T_i} \right) \right )} \exp{\left (
- \frac{t_i}{t} \right )}} & {\rm for} \ \ t < T_i \\
~ & ~ \\
\displaystyle{F_i \left ( \frac{t}{T_i} \right )^{-\alpha_i}
\exp{\left ( - \frac{t_i}{t} \right )}} & {\rm for} \ \ t \ge T_i \\
\end{array}
\right .
\label{eq: fc}
\end{equation}

\noindent for both the prompt (index `i=\textit{p}') $\gamma$\,-\,ray and initial X-ray decay and for the afterglow (`i=\textit{a}'),
modeled so that the complete light curve $f_{tot}(t) = f_p(t) + f_a(t)$ contains two sets of four free parameters $(T_{i},F_{i},
\alpha_i,t_i)$. The transition from the exponential to the power law (PL) occurs at the point $(T_{i},F_{i}e^{-t_i/T_i})$, where the
two functional forms have the same value. The parameter $\alpha_{i}$ is the temporal PL decay index, and the time $t_{i}$ is the
initial rise timescale. We exclude the cases when the fitting procedure fails or when the determination of $1 \sigma$ confidence
intervals does not fulfill the Avni (1976) $\chi^{2}$ prescriptions; see the XSPEC manual \footnote{http://heasarc.nasa.gov/xanadu/xspec/manual/XspecSpectralFitting.html}. Thus, we ended up with a sample of 183
GRBs.  We compute the source rest-frame isotropic
luminosity $L_a$ in units of erg $s^{-1}$ in the {\it Swift} XRT bandpass, $(E_{min}, E_{max})=(0.3,10)$ keV as follows:

\begin{equation}
L_a= 4 \pi D_L^2(z) \, F_X (E_{min},E_{max},T_a) \cdot \textit{K} ,
\label{eq: la}
\end{equation}

where $D_L(z)$ is the luminosity distance for the redshift $z$, assuming a flat $\Lambda$CDM cosmological model with $\Omega_M=0.3$ and $H_0=70$ km $s^{-1} Mpc^{-1}$, $F_X$ is the measured X-ray energy flux in erg $cm^{-2} s^{-1}$, and \textit{K} is the \textit{K}-correction for cosmic expansion. For {\it Swift} GRBs, the \textit{K}-correction is simply $(1+z)^{(\beta-1)}$, where $\beta$ is the X-ray spectral index of the plateau phase. For \textit{Fermi} GRBs, the \textit{K}-correction is the solution of this integral:

\begin{equation}
\frac{\int_{\frac{100}{(1+z)}}^{\frac{1000}{(1+z)}}N(E)}{{\int_{_{100}}^{_{1000}}N(E)}},
\end{equation}

where $N(E)$ is the functional form of the spectrum, represented by either a CPL or a Band function. We here note that the luminosity is calculated for both $L_{peak}$ and $L_a$ in a consistent rest-frame band, which is the {\it Swift}-BAT, XRT and Fermi-GBM bands, for each GRB and hence such a luminosity computation does not lead to any induced correlation. We downloaded the light curves from the {\it Swift} web page repository, \footnote{http://www.swift.ac.uk/burst\texttt{\_}analyser} and we derived the spectral parameters following Evans et al. (2009). As shown in Dainotti et al. (2010), requiring an observationally homogeneous sample in terms of $T^{*}_{90}$ and spectral lag properties implies removing
short GRBs ($T_{90} \leq 2$ s) and SEE from the analysis. We separated the GRBs cataloged as SEE in Norris \& Bonnel (2006),
Levan et al. (2007), Norris et al. (2010). For the evaluation of the remaining SEE GRBs we follow the definition of Norris et al. (2010),
who identify SEE events as those presenting short spikes followed, within $10$ s, by a decrease in the intensity emission by a factor
of $10^{3}-10^{2}$, but with almost negligible spectral lag. Moreover, because there are long GRBs for which an associated SN has
not been detected, such as, for example the nearby $z=0.09$ SN-less GRB 060505, the existence of a new group of long GRBs without SNe
has been suggested, thus highlighting the possibility of two types of long GRBs, with and without SNe. Therefore, in the interest of
selecting an observational homogeneous class of objects, we segregate long GRBs with no associated SNe from the other categories. Under
this specific criterion, all the GRB-SNe that follow the Hjorth \& Bloom (2011) classification are considered separately. Within the
GRB-SN sample, we applied a further classification, which is an update of the one of Hjorth \& Bloom (2011). This identifies GRB-SNe
subsamples based on the quality of the identification of SNe associated with the GRB. The categories considered are: A) strong
spectroscopic evidence for an SN associated with the GRB, B) a clear light curve bump as well as some spectroscopic evidence suggesting
the LONG-SNe association, C) a clear bump on the light curve consistent with the GRB-SN associations, but no spectroscopic evidence of
the SN, D) a significant bump on the light curve, but the inferred SN properties are not fully consistent with other GRB-SNe associations
or the bump is not well sampled or there is no spectroscopic redshift of the GRB; E) a bump, either of low significance or inconsistent
with other observed GRB-SNe identifications, but with a spectroscopic redshift of the GRB. Similarly, to evaluate samples that are
homogeneous regarding the ratio between $\gamma$-ray and X-ray fluence, we separated all XRFs from the other mentioned categories.
The selection criteria are applied in the observer frame.

In all that follows, $L_{peak}$ ($erg$ $s^{-1}$) is defined as the prompt emission peak luminosity over a $1$ s interval. Following Schaefer
(2007) we compute $L_{peak}$ as follows:

\begin{equation}
L_{peak}= 4 \pi D_L^2(z) \, F_{peak} (E_{min},E_{max},T_a) \cdot \textit{K} ,
\label{eq: lx}
\end{equation}

\noindent where $F_{peak}$ is the measured gamma-ray energy flux over a $1$ s interval ($erg$ $cm^{-2} s^{-1}$). To further create a sample with more homogeneous spectral features, we consider only the GRBs for which the spectrum computed at $1$ s has a smaller $\chi^2$ for a single PL fit than for a cutoff power law (CPL). Specifically, following Sakamoto et al. (2011), when the $\chi_{CPL}^{2}-\chi_{PL}^{2}<6$, the PL fit is preferred. In addition, for all GRBs that satisfy this criterion there is not a substantial difference in spectral fitting results if one considers either a PL or a CPL. 

As we have already anticipated in the introduction, the plane is confirmed also for GRBs observed by the {\it Fermi}-GBM. We consider a subsample of 76 GRBs, which are observed to have a plateau and are detected by both the {\it Fermi}/GBM and {\it Swift}. Of these 76 GRBs, we have selected 47 GRBs using the following selection criteria: $\delta_{F_{peak}}/F_{peak} \leq 1$, $\delta_\alpha/\alpha \leq 1$, $\delta_\beta/\beta \leq 1$ (where $\alpha$ and $\beta$ are the spectral parameters for the high-energy and the lower-energy tail for the Band function). This guarantees not only that the errors in the determined parameters are smaller than the parameters themselves, but also that these results are robust and independent of the instrument selected to measure the peak flux in the prompt emission. Among these 47 GRBs, $34$ are long, $13$ are gold, $3$ are SSE, $5$ are GRB-SNe associated and $5$ are XRFs.
%Indeed, we computed $L_{peak}$ from $F_{peak}$ at 1024 ms. 
We note that $L_{peak}$ is computed from $F_{peak}$ which has been binned at 1024 ms. 
The choice of the prompt peak luminosity as a third parameter guarantees that there is no contamination from the early afterglow, and thus the correlation with the afterglow luminosity is intrinsic. A confirmation of this statement comes from the use of peak luminosity computed in the $10-1000$ keV {\it Fermi} band, a much larger energy range compared to the {\it Swift}-BAT range ($15-150$ keV). For the GRBs observed by {\it Swift} we additionally discard six GRBs that were better fitted with a blackbody model than with a PL. This full set of requirements reduces the sample to $132$ long GRBs. Finally, we construct a subsample by including strict data quality and the following morphology requirements: the beginning of the plateau should have at least five data points, and the plateau should not be too steep (the angle of the plateau must be less than $41\degree$)\footnote{The angle of the plateau is obtained using trigonometry as the difference between the fluxes, $\Delta_F=F_i-F_a$, where $i$ is the time of the beginning of the plateau emission divided by the difference between $\delta_T=T_a-T_i$.}. The first of the above selection rules guarantees that the light curves clearly present the transition from the steep decay after the prompt emission to the plateau
phase. The number of points required for the W07 fit should be at least four, since there are four free parameters in the model,
one of which should be after the end of the plateau. Thus, the requirement of six points in total (five at the start and at least
one after the plateau) ensures a minimum number of points to have both a clear transition to the plateau phase and simultaneously
to constrain the plateau. This data quality criterion defines the gold sample, which includes $45$ GRBs. We have also confirmed through
the {\it T}-test that this gold sample is not statistically different in terms of ($L_a$, $T_a$, $L_{peak}$, $z$) distribution from
the full sample, thus showing that the choice of this sample does not introduce any biases, such as the Malmquist or Eddington
ones, against high luminosity and/or high redshift GRBs. Specifically, $L_a$,$T_a$,$L_{peak}$, and $z$ for the gold sample present
similar Gaussian distributions, but with smaller tails than the total sample (see Dainotti et al. 2015a); thus, there is no shift
of the distribution toward high-luminosities, larger times, or high-redshift. Hence, the selection cut naturally removes the majority
of the high error outliers of the variables involved, thus reducing the scatter of the relation for the gold sample.

We analyzed separately from the gold sample the following GRB categories: SEE (Norris \& Bonnel 2006; Levan et al. 2007;
Norris et al. 2010), the complete GRB-SNe sample (Hjorth \& Bloom 2011), the subsample of GRBs spectroscopically associated
with SNe (classifications A, B, and C from Hjorth \& Bloom 2011), XRFs, and long GRBs excluded from the ultra-long GRB category
and the previous categories. 

\section{The 3D Relation for Long GRBs, XRF, SEE, GRB-SNe, and the Gold Sample}\label{3D correlation}

\begin{figure}
\includegraphics[width=0.50\hsize,height=0.50\textwidth,angle=0,clip]{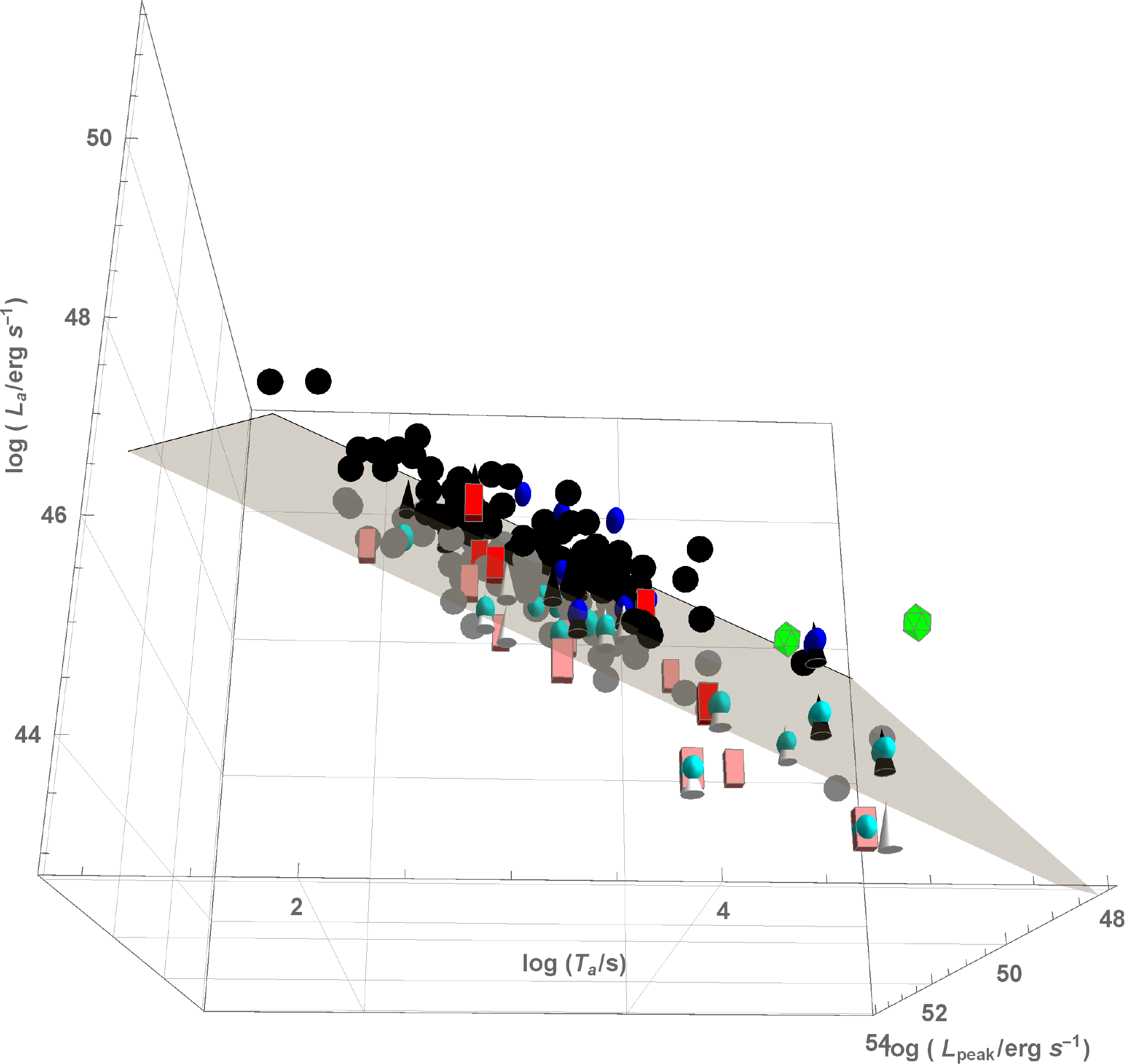}
\includegraphics[width=0.50\hsize,height=0.50\textwidth,angle=0,clip]{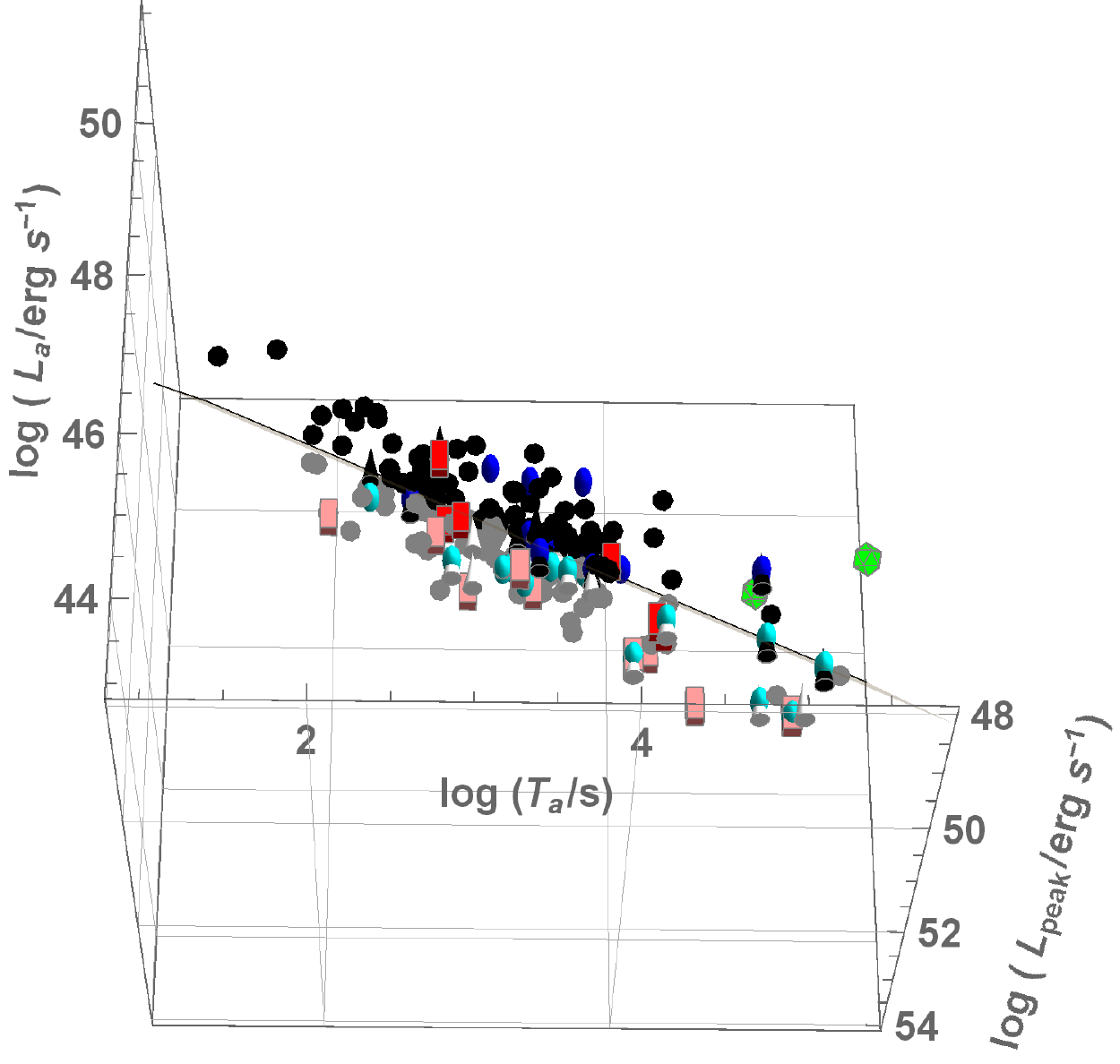}
\caption{Left panel: 183 GRBs in the $(L_{a}, T_{a}, L_{peak})$ space, with a plane fitted using the 183 GRBs, including GRB-SNe (cones),
  XRFs (spheres), SEE (cuboids), long GRBs (circles), and ultra-long GRBs (polyhedrons). Darker colors indicate data
  points above the plane, while lighter colors indicate data points below the plane. This figure shows one of the possible projections. Right panel: the same data are shown, but for an edge-on projection.}
\end{figure}

Figure (1) shows all 183 GRBs in the $(L_{a}, T_{a}, L_{peak})$ parameter space, divided into five categories: GRB-SNe (cones), XRFs (spheres), SEE (cuboids), long GRBs (circles), and ultra-long GRBs (polyhedrons). Darker colors indicate data points above the plane, while lighter colors indicate data points below the plane. It can be noted that the separate subclasses of GRBs show greater spread about the plane than the long GRB sample. Using the method described in Dainotti et al. (2016), we use the parameters
$L_{a}$, $T_{a}$, and $L_{peak}$ to create a best-fit plane for the GRB categories. When we parameterize this plane using the angles
$\theta$ and $\phi$ of its unit normal vector, the following formula is used: 

\begin{equation}
\log L_a = C_o - \cos(\phi) \tan(\theta) \log T_a - \sin(\phi) \tan(\theta)\log L_{peak}
\end{equation}

\noindent where $C_o = C(\theta,\phi,\sigma_{int})+z_o$ is the normalization of the plane correlated with the other variables, $\theta$,
$\phi$, and $\sigma_{int}$; while $z_o$ is an uncorrelated fitting parameter related to the normalization and $C$ is the covariance function.
For simplicity, we will rewrite the previous formula in the following way:
\begin{equation}
\log L_a = C_o + a \times \log T_a + b \times \log L_{peak}
\end{equation}

\noindent where $a(\theta,\phi)=-\cos(\phi) \tan(\theta)$, and $b(\theta,\phi)=-\sin(\phi) \tan(\theta)$. This normalization of the
plane allows the resulting parameter set, $\theta$, $\phi$, $\sigma_{int}$, and $z_o$ to be uncorrelated and provides explicit error
propagation. For example, with the updated gold sample, we obtain a new optimal plane:

\begin{equation}
\log L_{a}=(17.65 \pm 5.7) - (0.83 \pm 0.10) \log T_{a} + (0.64 \pm 0.11)\log L_{peak},
\end{equation}

\noindent where $C_o=(17.65 \pm 5.7)$, $a=-(0.83 \pm 0.10)$ and $b=(0.64 \pm 0.11)$. All of the fits shown in this paper were performed using
the D'Agostini method \citep{Dago05}. Uncertainties are always given as $1 \sigma$.

For the updated gold plane $\sigma_{int}=0.316 \pm 0.039$, which is within $1 \sigma$ of the previously obtained value of  $\sigma_{int}=0.27 \pm 0.04$. The {\it $R_{adj}^2$}
for the gold sample has slightly increased from $0.8$ to $0.81$, but remains comparable to the original gold sample. {\it $R_{adj}^2$} gives a version of the coefficient of determination, {\it $R^2$}, which is adjusted for the number of parameters in the model. The Pearson correlation coefficient, $r$, is $0.90$ with a probability of the same sample occurring by chance, $P=1.75 \times 10^{-17}$. The normalization of the plane, $C(\sigma_{int}, \phi, \theta)$, is given by

\begin{equation}
C = 13.90  - 62.28 \theta^2 - 0.29 \sigma_{int} +  0.38 \sigma_{int}^2 - 8.23 \phi - 0.05 \sigma_{int} \phi + 
 15.13 \phi^2 + \theta (99.62 - 0.10 \sigma_{int} +  90.31 \phi).
\end{equation}

\begin{table}[H]
\centering
\begin{tabular}{| l | l | l | l | l | l |}
\hline
	Category & $C_o$& {\it a} & {\it b} & $\sigma_{int}$ & {\it N} \\ \hline
	Gold & 17.65 $\pm$ 5.68 & -0.83 $\pm$ 0.1 & 0.64 $\pm$ 0.11 & 0.32 $\pm$ 0.04 & 45 \\ \hline
	SNe ABC & 20.87 $\pm$ 6.5 & -1.03 $\pm$ 0.12 & 0.58 $\pm$ 0.13 & 0.33 $\pm$ 0.08 & 11 \\ \hline
	SEE & 14.11 $\pm$ 8.16 & -1.05 $\pm$ 0.14 & 0.71 $\pm$ 0.16 & 0.39 $\pm$ 0.09 & 15 \\ \hline
	Long & 14.52 $\pm$ 3.67 & -0.87 $\pm$ 0.06 & 0.7 $\pm$ 0.07 & 0.41 $\pm$ 0.03 & 132 \\ \hline
	SNe Total & 10.19 $\pm$ 6.55 & -0.78 $\pm$ 0.12 & 0.77 $\pm$ 0.13 & 0.5 $\pm$ 0.08 & 22 \\ \hline
	XRF & 9.03 $\pm$ 7.14 & -0.71 $\pm$ 0.14 & 0.79 $\pm$ 0.13 & 0.53 $\pm$ 0.08 & 27 \\ \hline
\end{tabular}
\caption{Table of best-fit values for relation plane parameters in order of increasing scatter, $\sigma_{int}$.}
\end{table}

\noindent The correlation was also calculated for all of the mentioned GRB subclasses. 
The values for these fits are shown in Table (1) which shows subsamples in order of increasing scatter, $\sigma_{int}$. The panels of Figures (2) and (3) show the
fitted plane in projection for all mentioned subclasses. As we can see, going from the left to the right in Figure (2), the
scatter decreases from the XRF to the long sample, and it further decreases when we consider the planes shown in Figure (3) going
from SEE to GRB-SNe ABC, and finally reaching the smallest scatter in the right panel with the gold sample. These planes also show that the GRB-SNe ABC category, which is strongly
associated with SNe, is better correlated than that of the total GRB-SNe sample. This confirms a previous study performed in a 2D parameter space using the $L_a-T^{*}_a$ correlation.

%\begin{table}
%\centering
%\begin{tabular}{ | l | l | l | l | }
%\hline
%	Categories & $R_{adj}^2$ & R & Probability \\ \hline
%	Gold & 0.81 & 0.9 & 1.75E-17 \\ \hline
%	SNe ABC & 0.95 & 0.98 & 2.91E-7 \\ \hline
%	SEE & 0.87 & 0.94 & 1.98E-7 \\ \hline
%	Long & 0.76 & 0.87 & 1.99E-42 \\ \hline
%	SNe & 0.89 & 0.95 & 3.09E-11 \\ \hline
%	XRF & 0.86 & 0.93 & 1.47E-12 \\ \hline
%\end{tabular}
%\caption{Statistical parameters for a linear-fit plane of each category}
%\end{table}

\begin{figure}[H]
\includegraphics[width=0.32\hsize,height=0.32\textwidth,angle=0,clip]{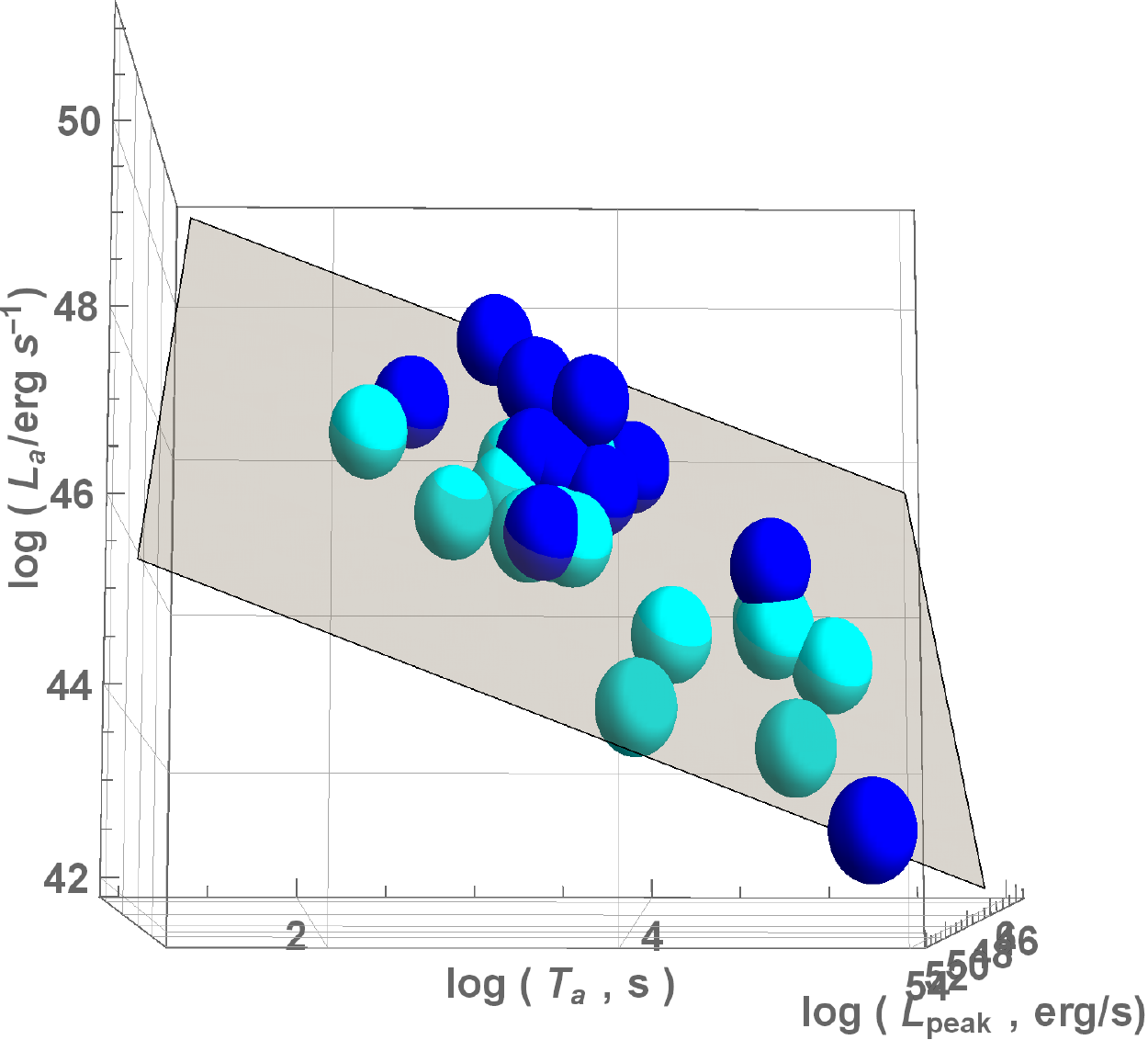}
\includegraphics[width=0.32\hsize,height=0.32\textwidth,angle=0,clip]{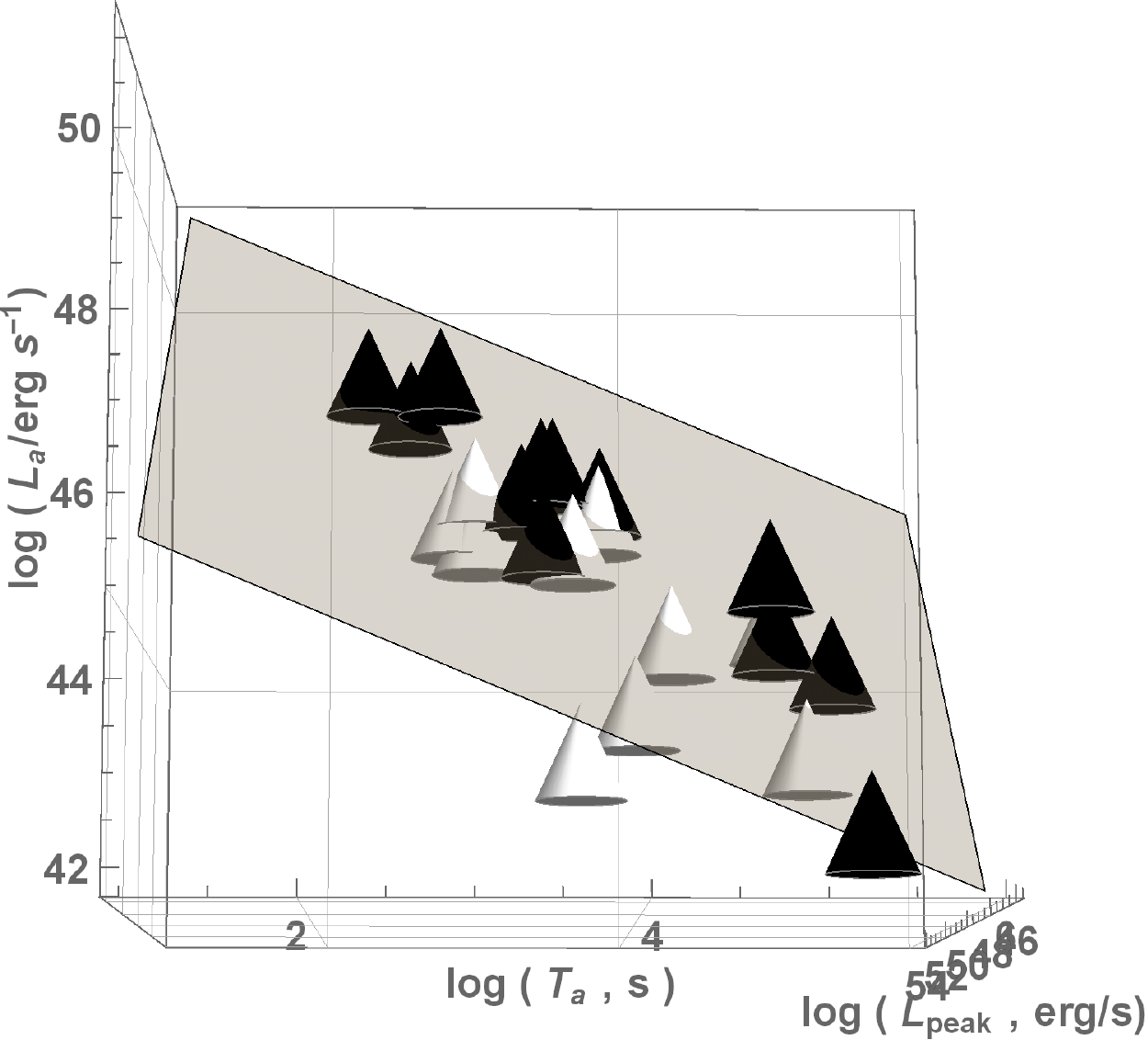}
\includegraphics[width=0.32\hsize,height=0.32\textwidth,angle=0,clip]{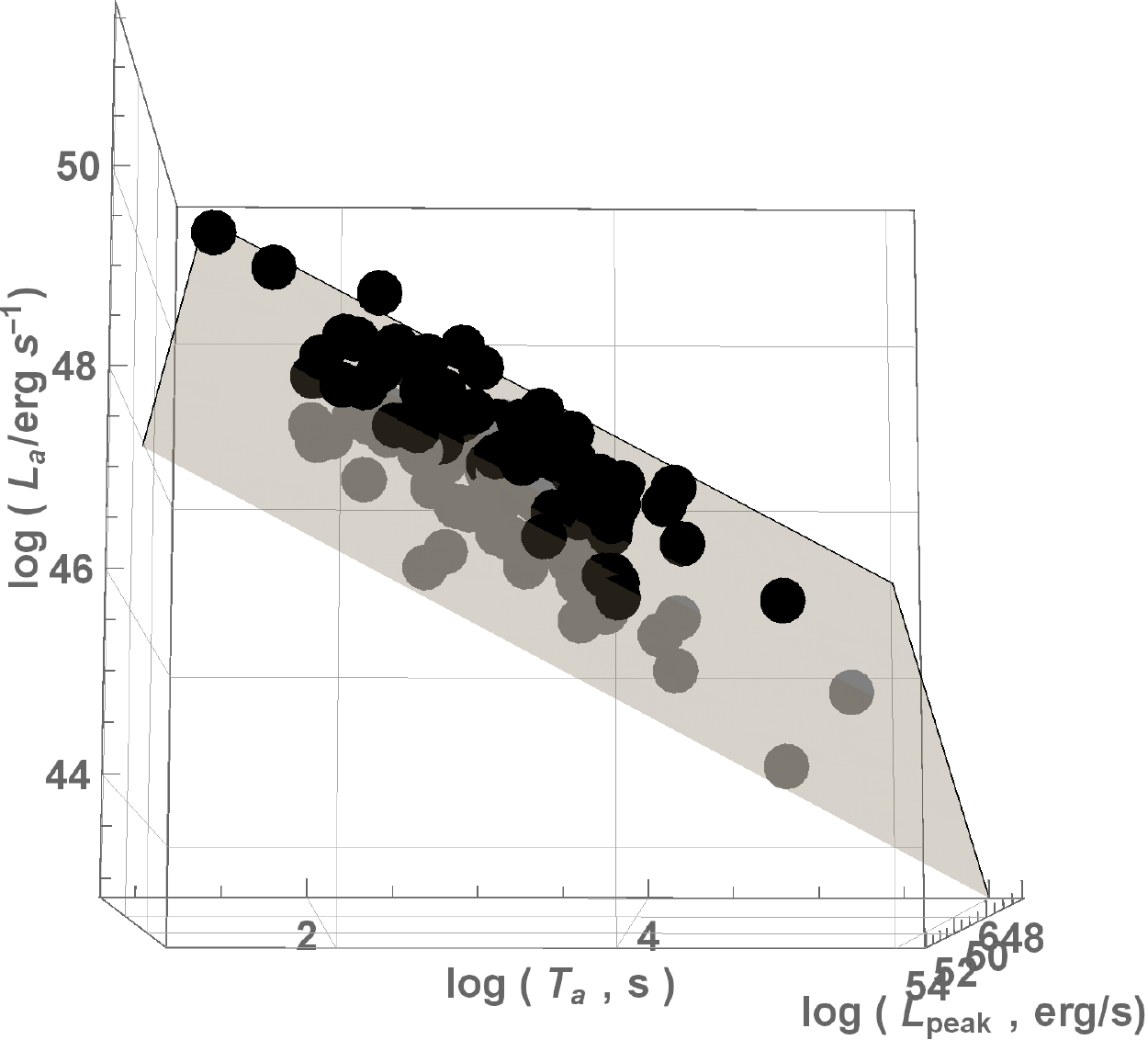}
\caption{Projection of the ($L_a,T_a,L_{peak}$) relation, in order of decreasing intrinsic scatter, for XRF, GRBs associated with SNe,
  and long GRBs respectively.}
\end{figure}

\begin{figure}[H]
\includegraphics[width=0.32\hsize,height=0.32\textwidth,angle=0,clip]{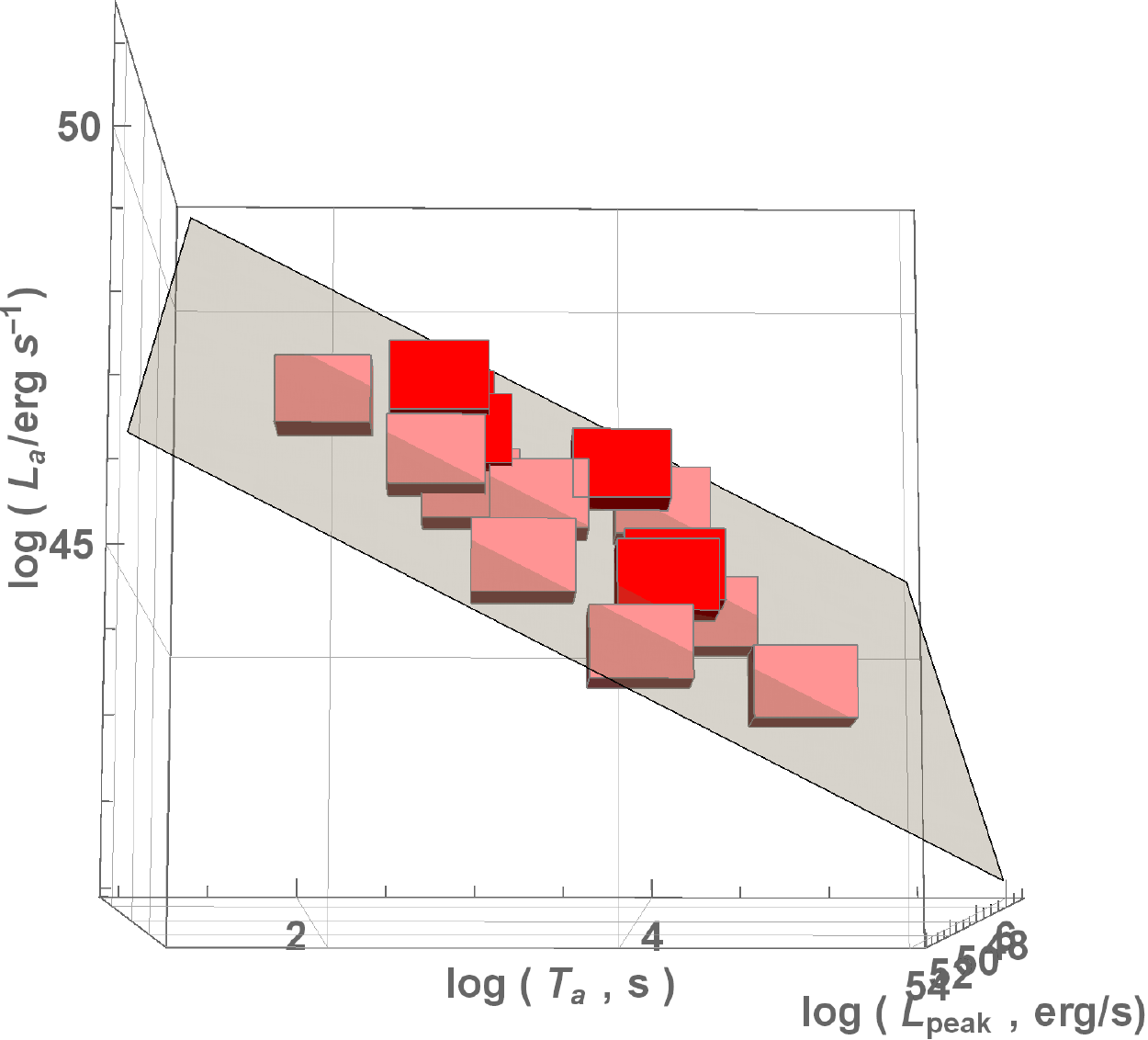}
\includegraphics[width=0.32\hsize,height=0.32\textwidth,angle=0,clip]{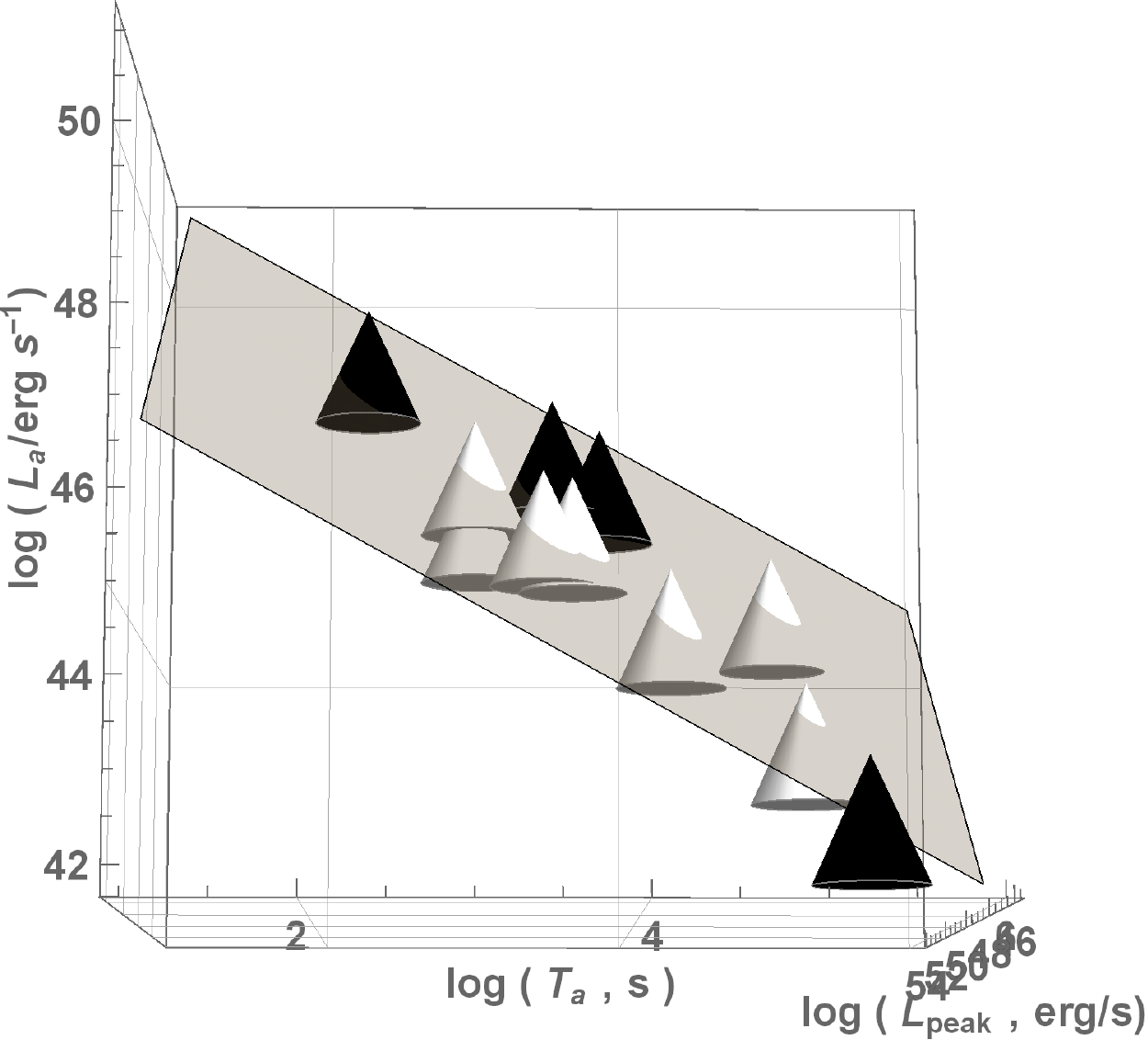}
\includegraphics[width=0.32\hsize,height=0.32\textwidth,angle=0,clip]{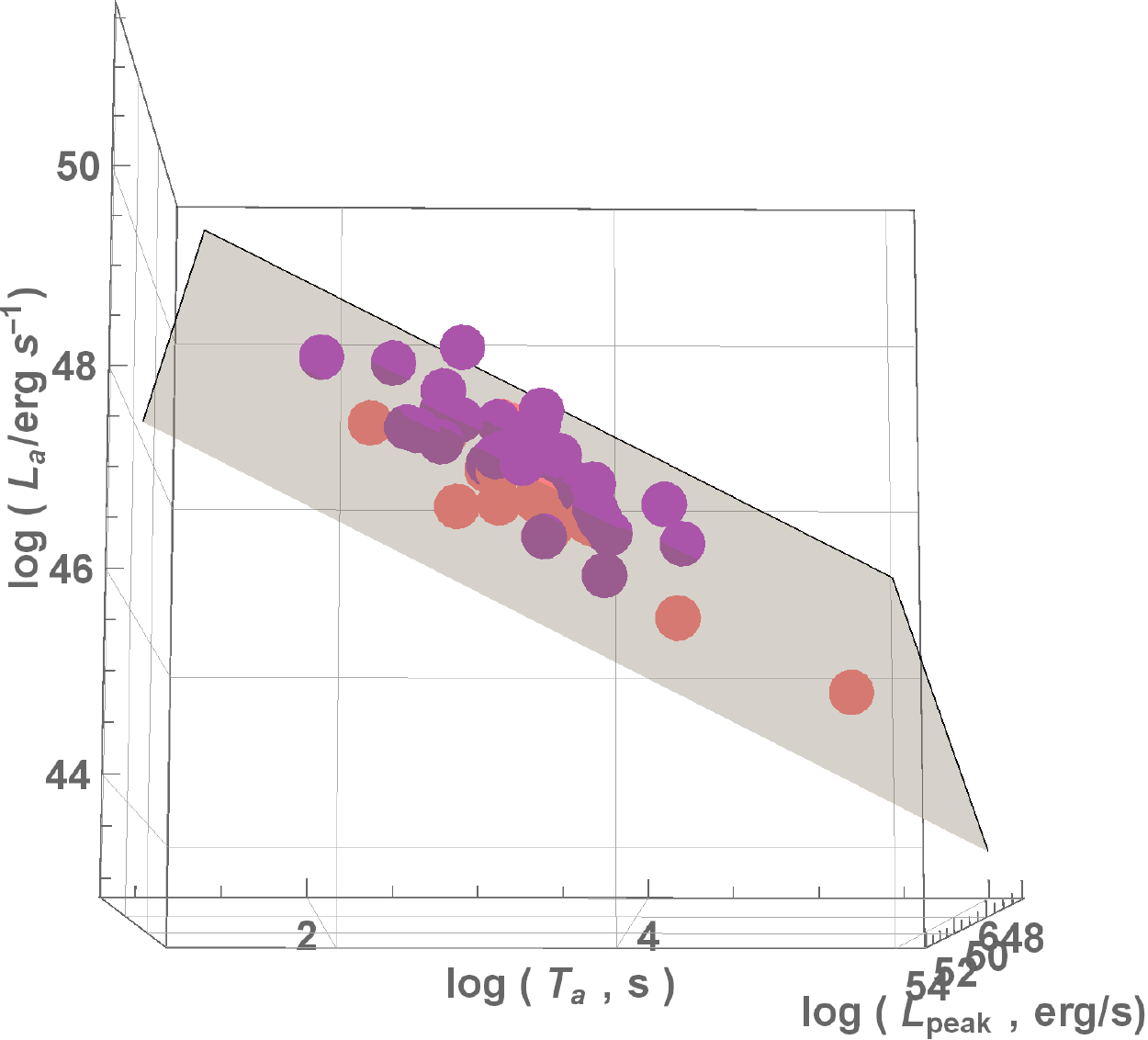}
\caption{Projection of the ($L_a,T_a,L_{peak}$) relation, in order of decreasing intrinsic scatter, for SEE, GRBs spectroscopically
  associated with SNe, and the gold sample, respectively.}
\end{figure}

\noindent 

There are a few key details to notice in Table (1). Crucially, the gold sample still has the lowest intrinsic scatter of all the fitted categories. Another significant feature is that all of the plane parameters are within $1 \sigma$ of the fundamental plane set by the gold sample. The planes of these categories are not statistically different; thus, we cannot hypothesize that these planes suggest different energy mechanisms, but we can conclude that the existence of the fundamental plane is confirmed to be driven by the gold sample features
rather than the category-based sampling. The one exception to the above is the case of the short GRBs with extended emission,
  the SEE category. Although from the distance to the plane of any particular GRB it is not possible to assign it to this category
  or another, the full SEE sample of 15 bursts has a mean distance to the gold plane of $0.56$ (see Fig. 4). The above implies a {\it z}-score of $-8.3$ for this sample, \footnote{The probability that a subsample of {\it n} members drawn from an underlying gaussian distribution having a dispersion of $\sigma$ will have a mean farther away from that of the underlying distribution by more than $X \sigma$, will be determined by the probability corresponding to a {\it z}-score of (sqrt({\it n})*X)/ $\sigma$.}, and hence
 a probability of 10$^{-5}$ that the same {\it z}-score test statistic could be obtained by chance for the same population, for details see Table 2. Table 2 gives the {\it z}-scores for all the sub classes treated, showing that the SEE type appears as a very clear outlyer. Thus, we can conclude that SEE GRBs are in all likelihood produced by a distinct physical mechanism. SEEs may be related to short bursts and hence come from a different progenitor.

\begin{table}[H]
\centering
\begin{tabular}{ | l | l | l | l | l | l | l|}
\hline
Category &  {\it z}-score & {\it N} \\ \hline
Gold & 0.0 & 45 \\ \hline
Long & -3.5 & 132 \\ \hline
SNe & -5.8 & 22 \\ \hline
XRF & -6.0 & 27 \\ \hline
SEE & -8.3 & 15 \\ \hline
\end{tabular}
\caption{Table of {\it z}-scores for each subsample.}
\end{table}

It can also be shown that no category distribution is significantly separated from the fundamental plane. In Figure (4), combined plots of the distribution of GRB geometric distance from the fundamental plane set by the gold sample are shown for each category. The fitting of the plane and the dispersion have been performed simultaneously to avoid fitting bias. Thus, the shift visible in Fig. 4 derives from the fact that the reference plane is the gold sample fundamental plane rather than each plane for each category. Indeed, the gold fundamental plane is placed in 0, as a reference plane. The center of the distributions for all of the GRB subcategories lie within 1$\sigma_{int}$ of the gold fundamental plane, again with the exception being the SEE class, which is seen to peak at a point where the gold sample has already fallen to very close to $0$. In addition, as it is visible from the right panel of Fig. 4, where the probability distributions of smoothed histograms are plotted, the distance of the peaks of the distribution between the gold sample and the SSE is the largest. For details about the definition and how the smoothed histogram has been computed, see Appendix 3.
This strengthens the possibility that the distance to the plane for the gold sample is a relevant discriminant between GRB categories. Of note is the fact that the ultra-long GRBs can be associated with the fundamental plane. One possibility that could explain this is discussed in Greiner et al. (2015), where it was found that an ultra-long GRB (GRB 111209A) was associated with an SN, which may indicate a magnetar origin. There are only two ultra-long GRBs in this sample, so a full analysis of this type will have to wait until a larger sample is available.

\begin{figure}[H]
\includegraphics[width=0.48\hsize,height=0.30\textwidth,angle=0,clip]{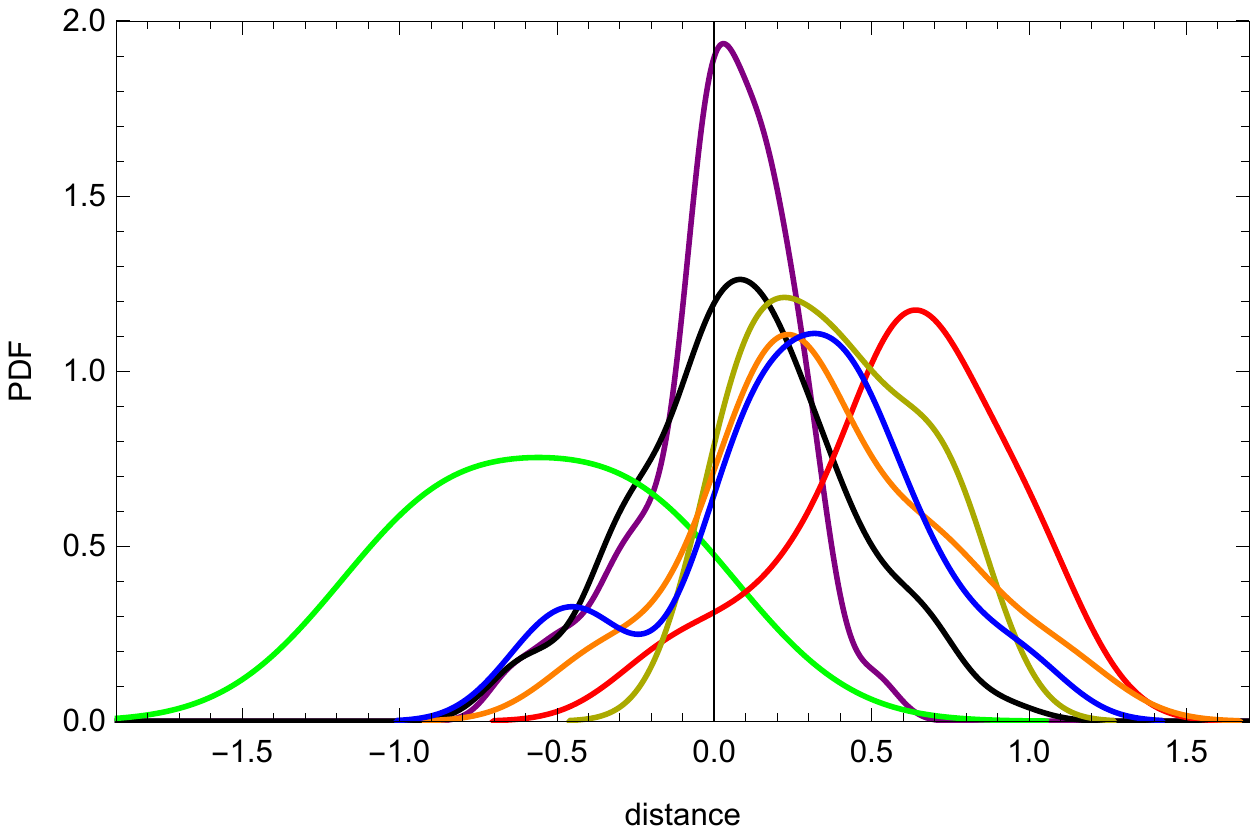}
\includegraphics[width=0.48\hsize,height=0.30\textwidth,angle=0,clip]{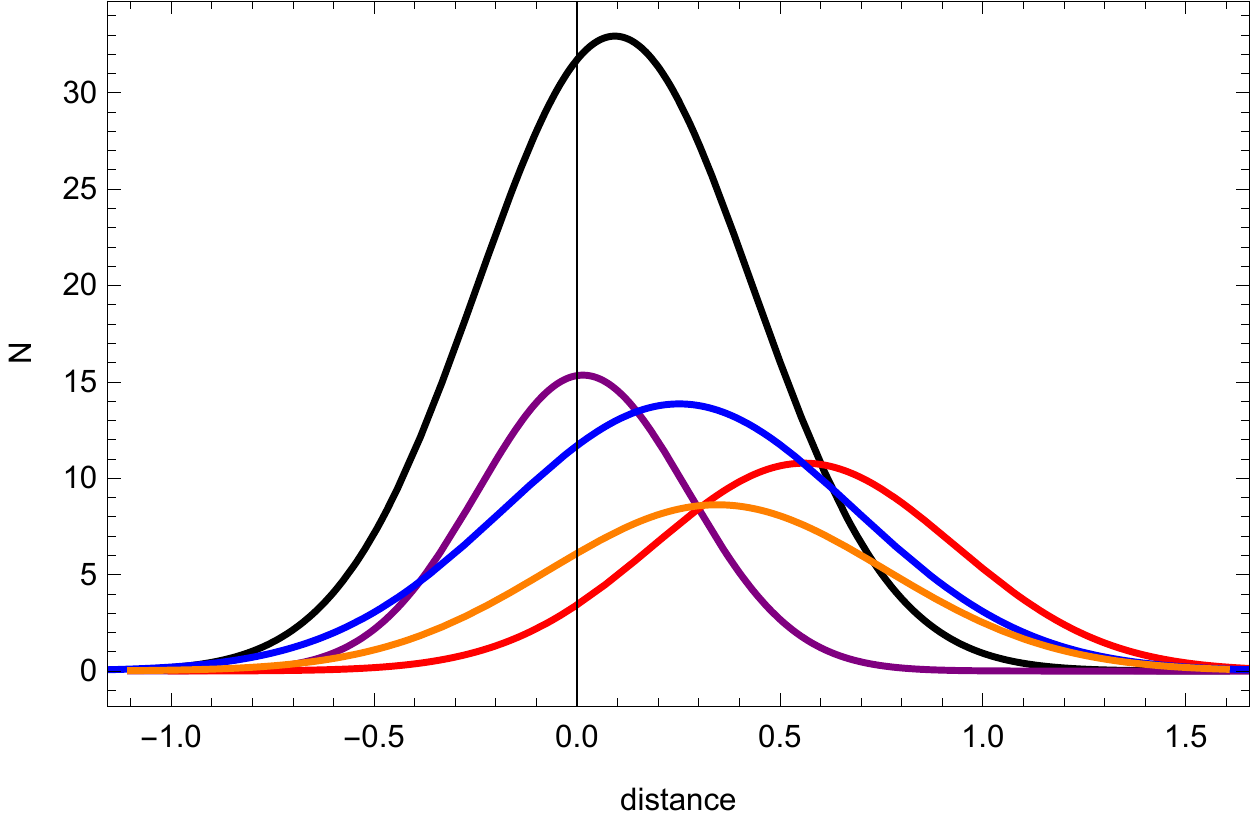}
\caption{Left: smoothed histogram showing the probability distribution function (PDF) of the distance distribution from the fundamental plane for GRBs of each category, including GRB-SNe (orange), GRBs spectroscopically associated with SNe (yellow), XRF (blue), SEE (red), gold sample GRBs (purple), long GRBs (black),
  and ultra-long GRBs (green). Right: A plot showing Gaussian fits of the distributions displayed in the left panel, using the same
  color scheme. A line perpendicular to $x=0$ is shown as the reference of the gold sample compared to the other categories. Ultra-long GRBs and GRBs
  spectroscopically associated with SNe are not shown, as they had too little data to reliably fit them.}
\end{figure}

In a further attempt to reduce the scatter of the correlation, we consider the rest-frame time $T^{*}_a$ as the dependent variable. This means that the length of the plateau depends on the peak luminosity and the luminosity at the end of the plateau itself. This choice is not only motivated from a mere hunt for a smaller scatter of the correlation, but also is dictated by the intrinsic physics, since the length of the plateau in the magnetar model scenario is determined by the luminosity at the same time. Thus, it can be derived from fundamental physics, under this particular model; see Rowlinson et al. (2014) for details about this derivation. We present the results of this fitting in Table 3. Analogous to Table 1, we report the various GRB categories. The equation for the plane in this case is written in the following way:

\begin{equation}
\log T_a = C^{'}_o + a^{'} \times \log L_a + b^{'} \times \log L_{peak}
\end{equation}

\begin{table}[H]
\centering
\begin{tabular}{ | l | l | l | l | l | l | l|}
\hline
Category & $C^{'}_o$& $a^{'}$ & $b^{'}$ & $\sigma^{'}_{int}$ & {\it N} & $\Delta_\sigma$ \\ \hline
SNe ABC & 21.13 $\pm$ 5.85 & -0.79 $\pm$ 0.15 & 0.38 $\pm$ 0.10 & 0.29 $\pm$ 0.07 & 11 & $12\%$ \\ \hline
Gold & 24.15 $\pm$ 6.79 & -0.75 $\pm$ 0.11 & 0.29 $\pm$ 0.10 & 0.30 $\pm$ 0.04 & 45 & $6\%$ \\ \hline
SEE & 16.73 $\pm$ 7.45 & -0.72 $\pm$ 0.15 & 0.39 $\pm$ 0.12 & 0.31 $\pm$ 0.08 & 15 & $21\%$\\ \hline
Long & 19.16 $\pm$ 4.18 & -0.69 $\pm$ 0.07 & 0.33 $\pm$ 0.06 & 0.38 $\pm$ 0.03 & 132 & $24\%$ \\ \hline
SNe Total & 18.35 $\pm$ 7.38 & -0.74 $\pm$ 0.17 & 0.39 $\pm$ 0.12 & 0.48 $\pm$ 0.08 & 22 & $4\%$ \\ \hline
XRF & 20.02 $\pm$ 8.23 & -0.58 $\pm$ 0.16 & 0.22 $\pm$ 0.12 & 0.48 $\pm$ 0.07 & 27 & $9\%$ \\ \hline
\end{tabular}
\caption{Table of best-fit values for relation plane parameters in order of increasing scatter, $\sigma_{int}$. Note. The last column $\Delta_\sigma$ indicates the difference in percentage about the computation using rest-frame time, $T_a$, as the dependent variable.}
\end{table}

We here note that the scatter of the relations considered with $T_a$ as a dependent variable is smaller than the ones with $L_a$, at least more than $4\%$ in the case of the GRB-SNe and reaching a $24\%$ reduction for the total long sample category, as it is indicated in the last column of Table $3$.
Thus, this new approach constitutes an improvement compared to the previous analysis, especially for the long GRBs. Further, as shown in Table 4 we see that the method is insensitive giving compatible results when using normalized or standardized variable.

\begin{table}[H]
\centering
\begin{tabular}{ | l | l | l | l | l | l | l|}
\hline
Category & $C_o'$& a$'$ & b$'$ & $\sigma_{int}$ & $\Delta_\sigma$ \\ \hline
Gold normalized & 21.91 $\pm$ 6.58 & -0.74 $\pm$ 0.10 & 0.29 $\pm$ 0.09 & 0.30 $\pm$ 0.04 & $6\%$ \\ \hline
Gold standardized & 21.08 $\pm$ 6.67 & -0.75 $\pm$ 0.11 & 0.29 $\pm$ 0.09 & 0.30 $\pm$ 0.04 & $6\%$ \\ \hline
\end{tabular}
\caption{Table of best-fit values for relation plane parameters using the gold sample where the dependent variable \textit{T}$_{a}$ is either normalized or standardized. Note. We note that there is not a significant change in the scatter, $\sigma_{int}$, from that of the gold sample in Table 3, showing that the fitting method used is not sensitive to scale differences.}
\end{table}

A reduction in the scatter might allow us to employ the 3D gold fundamental plane relation, in combination with other GRB relations, as cosmological tools. This may be possible based on a previous study of some of us (Cardone et al. 2009) that showed that adding the 2D Dainotti relation, $L_X-T^{*}_a$, to other five GRB relations reduces the resulting confidence intervals on the inferred distance moduli by $14\%$. The sample of the Dainotti relation used in Cardone et al. (2009) was composed of only $28$ GRBs versus the $45$ GRBs presented here. The $\sigma_{int}$ scatter of the 2D relation was $\sigma_{int}=0.33$ versus $\sigma_{int}=0.30$ of the current 3D relation. Thus, this increase in the sample size and $10\%$ decrease in the $\sigma_{int}$ possibly allow a reduction in the inferred cosmological parameters if we replace the 2D Dainotti relation with the current 3D gold fundamental plane, together with the other five GRB relations used in Schaefer et al. (2007).

We here note that the gold fundamental plane reaches a much smaller intrinsic scatter than the gold sample obtained with $L_{peak}$ computed using the full {\it Swift}-BAT GRBs. We have a reduction of the scatter of $27\%$, while the results of the long sample for the GBM are comparable within 1 $\sigma$ with the previous {\it Swift} results. This is a possible step forward in the use of the fundamental plane as a cosmological tool.

As previously noted in the introduction, the plane is confirmed also for GRBs observed by the {\it Fermi}-GBM. We have a sample of 47 GRBs that are in common among the sample observed by Swift and the sample observed by the GBM. 
We here present Table 5 that summarizes the results of the fitting for the long category using either the {\it Fermi} or {\it Swift} data by using $L_{peak}$ in erg $s^{-1}$. As shown in the table, the normalization coefficient found using {\it Fermi} data is larger than the normalization coefficient found using the {\it Swift} data. This is expected given GBM's larger energy band. We will not present the other categories, due to the paucity of the data. From the table of {\it Swift} and GBM we can see that using the same GRB set, but with a different spectral model for the prompt emission does not significantly alter results. As expected, we have a change in the normalization and consistent plane orientations confirming the physical nature of the 3D plane.

\begin{table}[H]
\centering
\begin{tabular}{| l | l | l | l | l | l |}
\hline
Category & $C_o$& $a$ & $b$ & $\sigma_{int}$ & N \\ \hline
Long (Fermi) & 21.34 $\pm$ 5.96 & -0.89 $\pm$ 0.07 & 0.58 $\pm$ 0.10 & 0.43 $\pm$ 0.07 & 34 \\ \hline
Long (Swift) & 17.22 $\pm$ 7.50 & -0.88 $\pm$ 0.09 & 0.65 $\pm$ 0.13 & 0.48 $\pm$ 0.07 & 34 \\ \hline
\end{tabular}
\caption{Table of best-fit values for relation plane parameters in order of increasing scatter, $\sigma_{int}$. These values are computed assuming $1024$ ms.} 
\end{table}

\section{Independence of the Fundamental Plane from Selected Parameters}\label{results}

\begin{figure}
\includegraphics[width=0.4\hsize,height=0.28\textwidth,angle=0,clip]{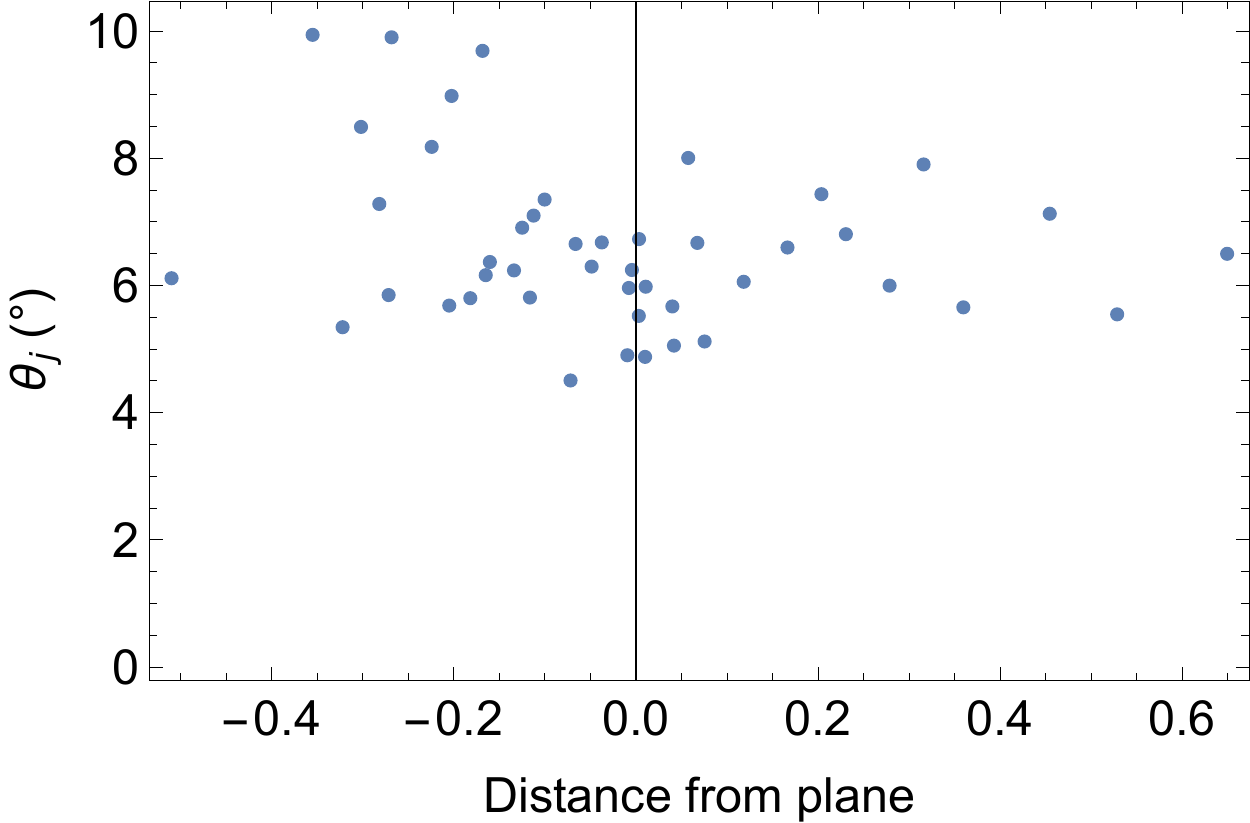}
\includegraphics[width=0.58\hsize,height=0.32\textwidth,angle=0,clip]{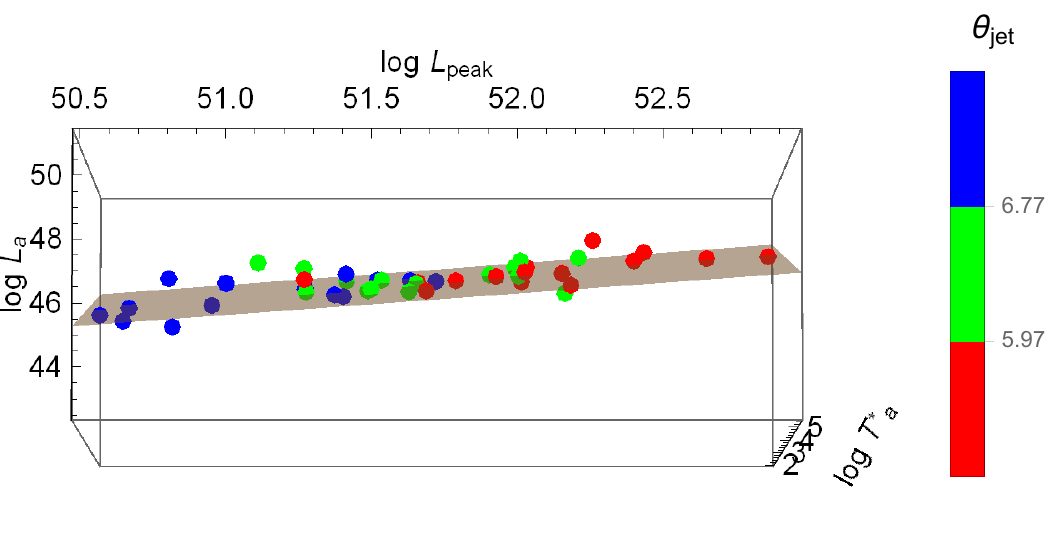}
\caption{Left: A 2D plot of $\theta_{jet}$ over distance from the fundamental plane for the gold sample. Right: color bar plot of fundamental plane with a color bar depending on $\theta_{jet}$.}
\end{figure}

\begin{figure}
\includegraphics[width=0.4\hsize,height=0.28\textwidth,angle=0,clip]{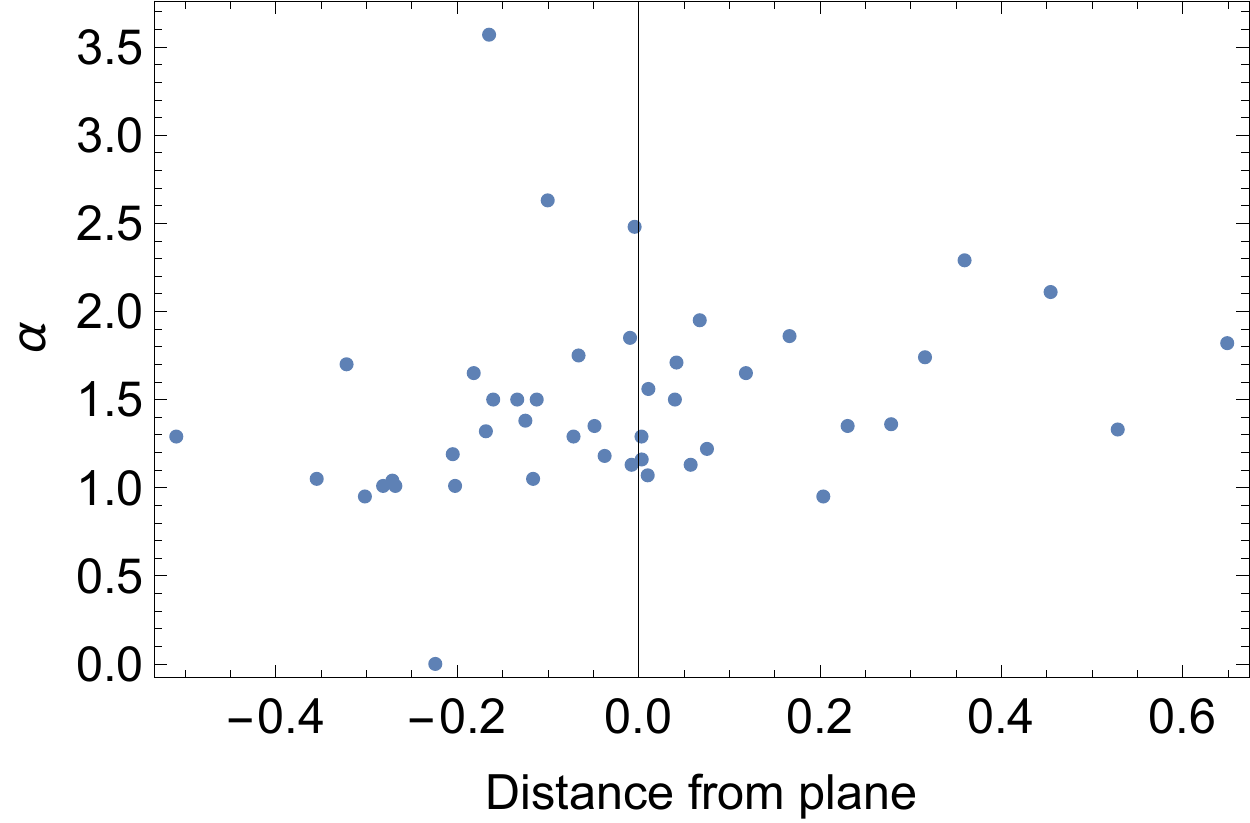}
\includegraphics[width=0.58\hsize,height=0.32\textwidth,angle=0,clip]{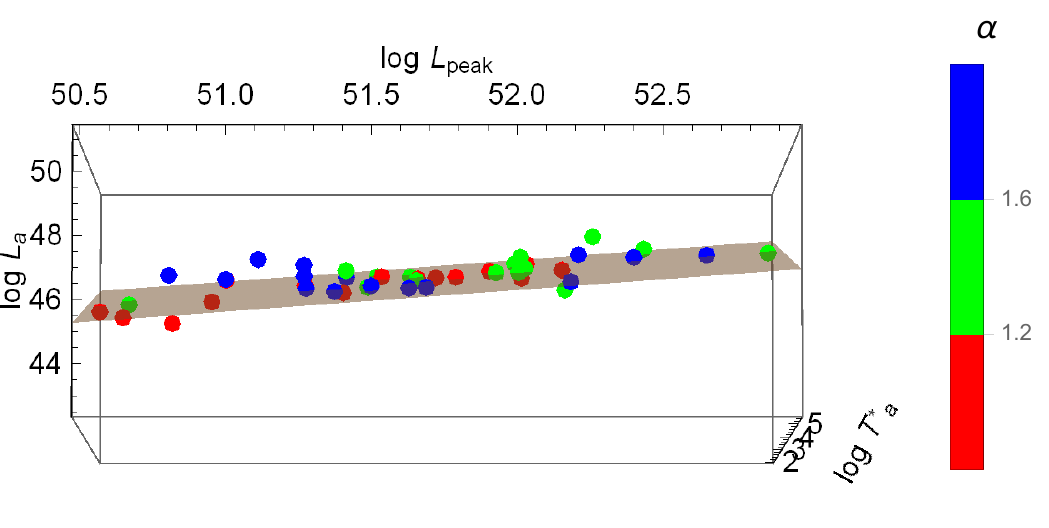}
\caption{Left: A 2D plot of $\alpha$ over distance from the fundamental plane for the gold sample. Right: color bar plot of fundamental
  plane with a color bar depending on $\alpha$.}
\end{figure}

\begin{figure}
\includegraphics[width=0.42\hsize,height=0.3\textwidth,angle=0,clip]{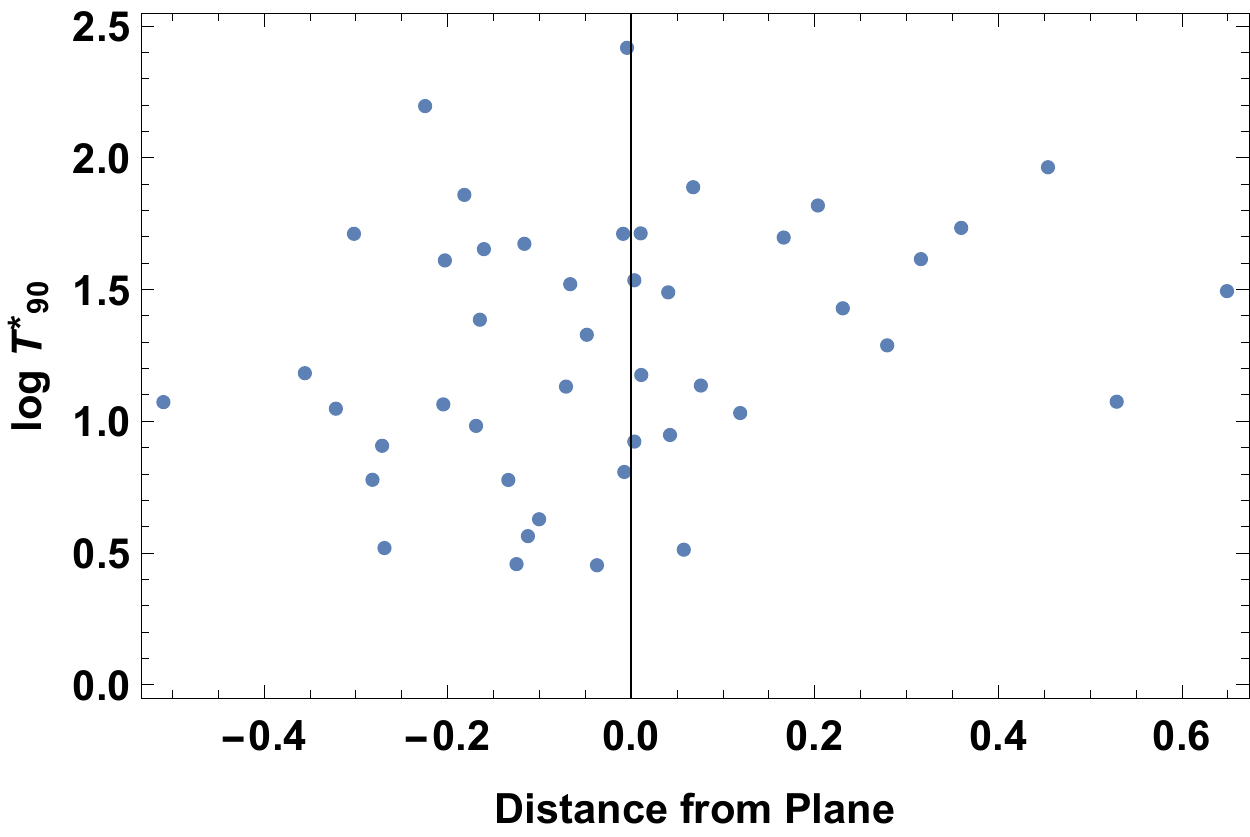}
\includegraphics[width=0.5\hsize,height=0.3\textwidth,angle=0,clip]{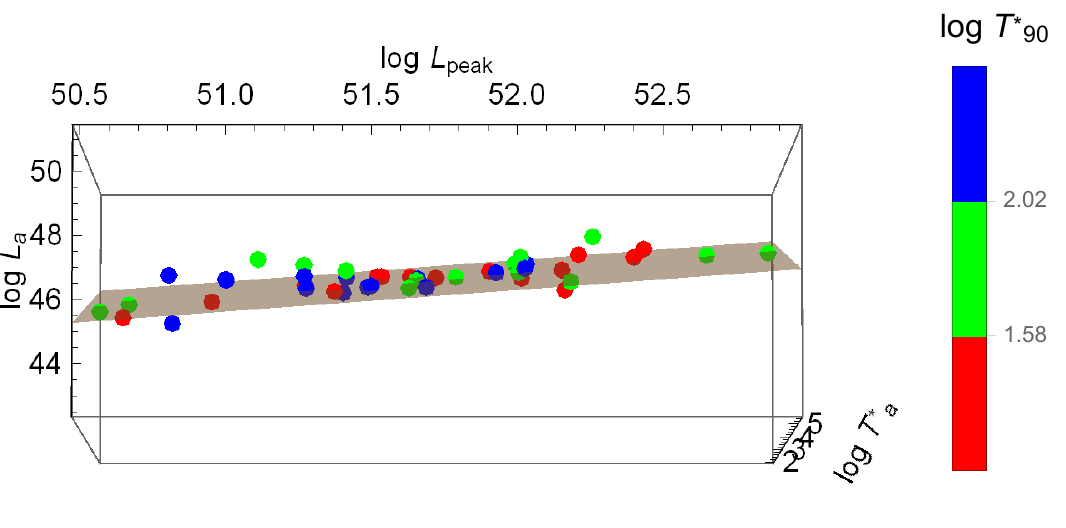}
\caption{Left: A 2D plot of $T^{*}_{90}$ over distance from the fundamental plane for the gold sample. Right: color bar plot of fundamental
  plane with a color bar depending on $T^{*}_{90}$.}
\end{figure}

\begin{figure}
\includegraphics[width=0.45\hsize,height=0.3\textwidth,angle=0,clip]{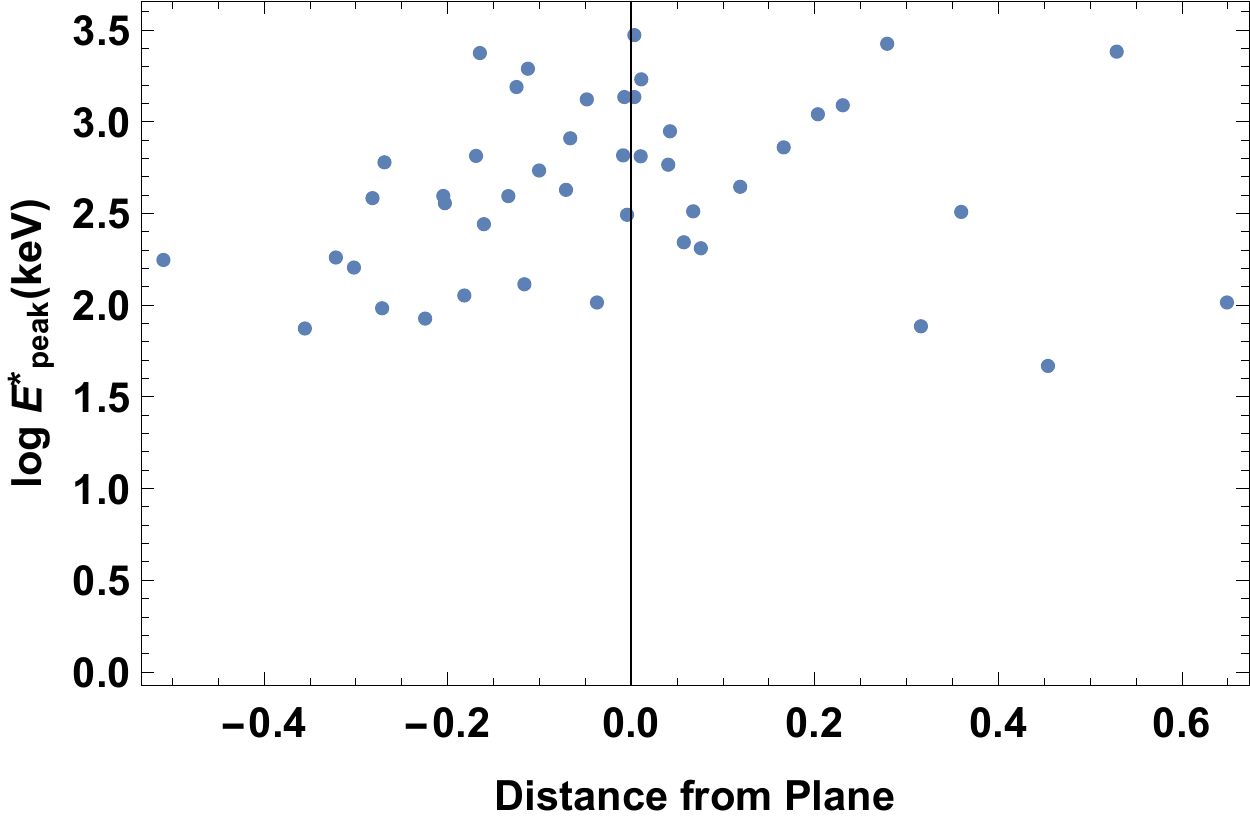}
\includegraphics[width=0.5\hsize,height=0.3\textwidth,angle=0,clip]{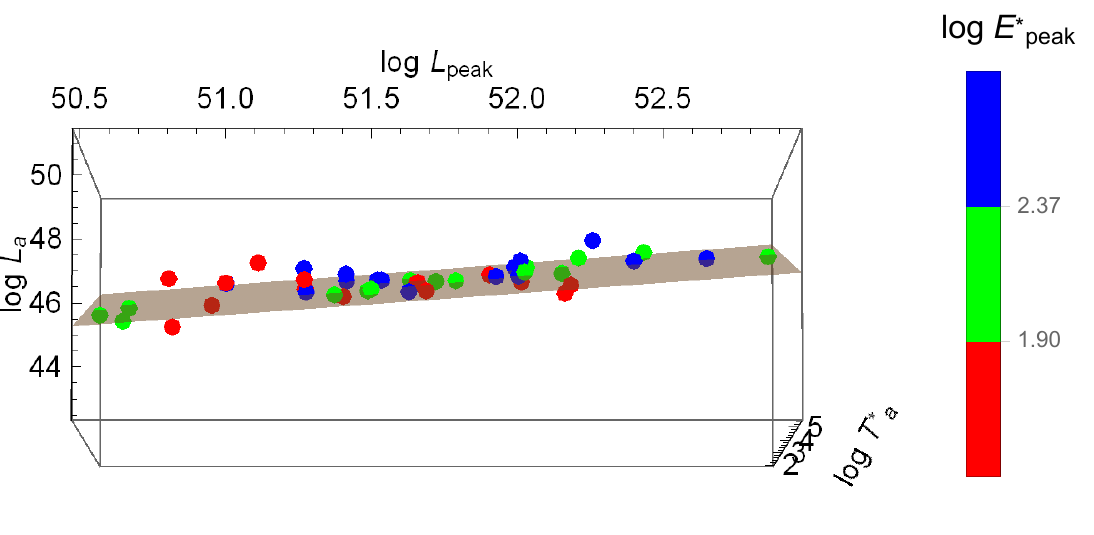}
\caption{Left: A 2D plot of $E^{*}_{peak}$ and the distance from the fundamental plane for the gold sample. Right: color bar plot of fundamental
  plane with a color bar depending on $E^{*}_{peak}$.}
\end{figure}

\begin{figure}
\includegraphics[width=0.49\hsize,height=0.36\textwidth,angle=0,clip]{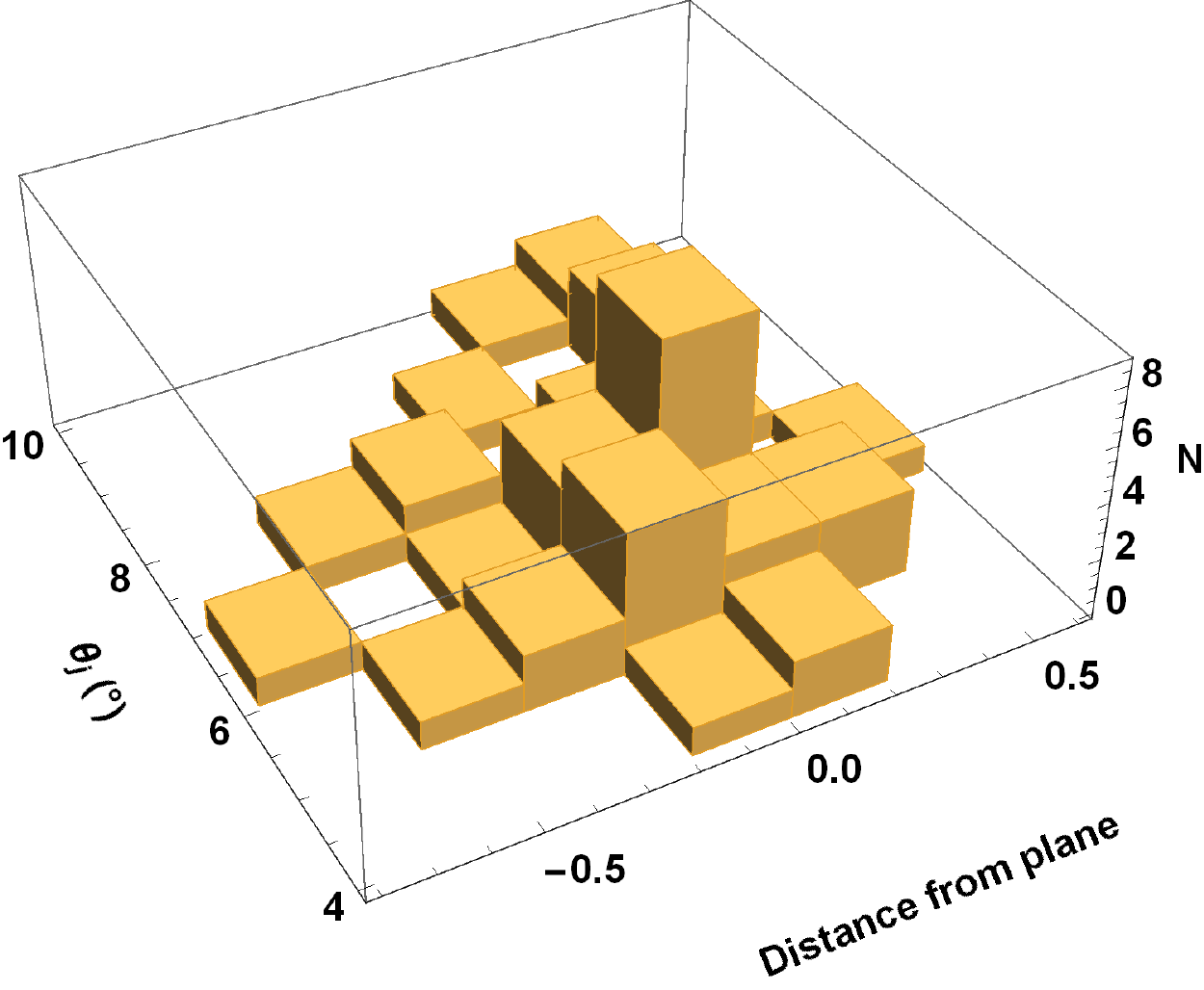}
\includegraphics[width=0.49\hsize,height=0.36\textwidth,angle=0,clip]{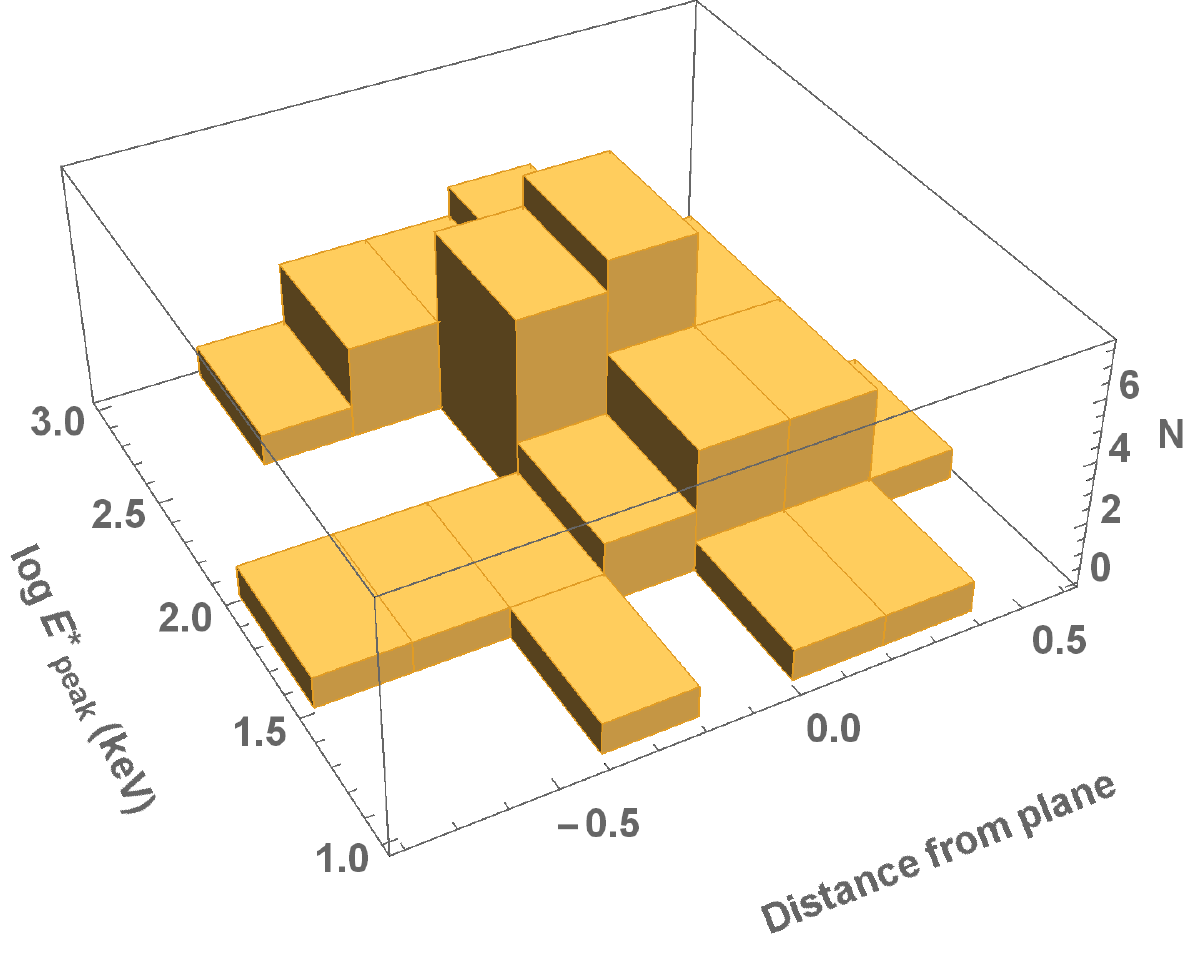}
\includegraphics[width=0.49\hsize,height=0.36\textwidth,angle=0,clip]{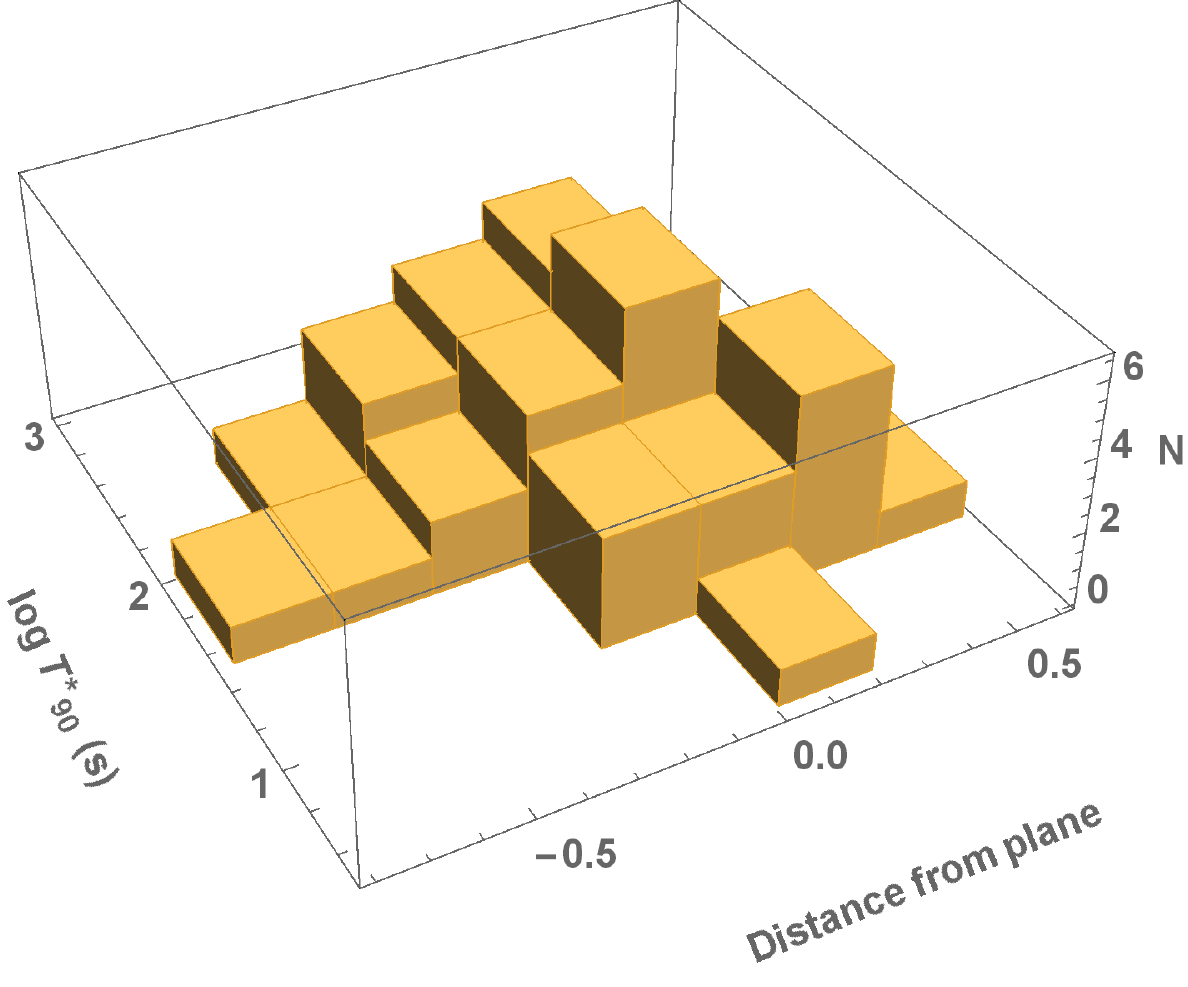}
\includegraphics[width=0.49\hsize,height=0.36\textwidth,angle=0,clip]{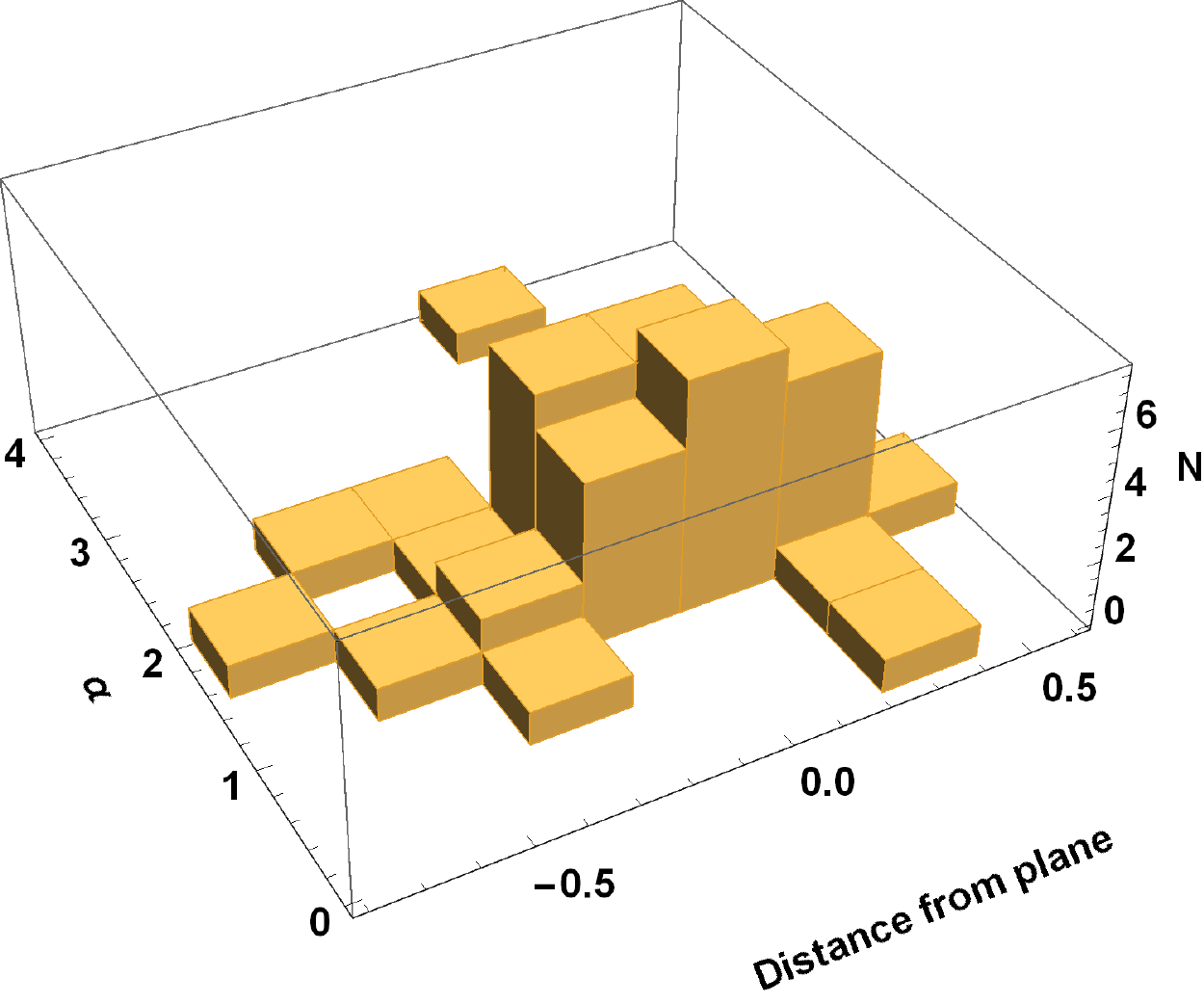}
\caption{3D histograms of the relationship between the same prompt and afterglow parameters of Figures (5)-(8) and the geometric distance
  from the fundamental plane.}
\end{figure}

With the goal of further reducing the scatter of this 3D relation, a number of independent GRB parameters were tested to see whether or not
a 4D relation existed that would significantly decrease the scatter of the fundamental plane. The tested quantities included $T^{*}_{90}$,
$E^{*}_{peak}$, $\alpha$, and $\theta_{jet}$, the first two rest-frame quantities.

We here check the dependence on the jet opening angle, but we will not discuss comparisons with several methods of computing this angle since it would be far beyond the scope of the paper. For details on how we estimate the jet opening angle, see the Appendix 1. 

In the left panels of Figures (5)-(8), there are plots of these parameters versus the geometric distance to the fundamental plane
derived from the gold sample. In the right panel of the same figures, the 4D color bar plots visually show the
relationship between the parameters and the positions of the points on the plane. If a correlation had existed between the parameters
and the plane, a clear pattern should have been seen in the scatter and color bar plots. There are no explicit trends in the scatter
plots or groupings of colors shown. The Spearman coefficient for these distributions is uniformly $\leq 0.40$, confirming the lack of
real correlation for these parameters. \footnote{The Spearman coefficient is a nonparametric measure of rank correlation. A Spearman
  coefficient less than $0.50$ generally means that the correlation is not meaningful.} 
We here note that since the Yonetoku relation is between $E_{peak}$ and $L_{peak}$ in the prompt emission, one could expect a correlation between $E_{peak}$ and the fundamental plane. However, this is a three-parameter correlation; thus while the contribution to this correlation is high for the presence of $L_{peak}$, it diminishes when we consider the correlation between $E_{peak}$ and ($L_a$,$T_a$). The Spearman and Pearson correlation coefficients for each of the parameters versus the geometric distance from the fundamental
plane are given in Table (6).

\begin{table}[H]
\centering
\begin{tabular}{ | l | l | l | l | l |}
\hline
	Parameter & Spearman $\rho$ & ${\it R_{adj}^2}$ & {\it R} & Probability\\ \hline
	$\alpha$       & 0.40 & 0.05 & 0.26 & 0.04 \\ \hline
	$\theta_{jet}$ & 0.21 & 0.03 & 0.23 & 0.06 \\ \hline
	$\log T^{*}_{90}$   & 0.26 & 0.02 & 0.22 & 0.08 \\ \hline
	$\log E^{*}_{peak}$ & 0.30 & 0.02 & 0.20 & 0.09 \\ \hline
\end{tabular}
\caption{Statistical and correlation parameters for linear fits of the parameters vs. geometric distance from the fundamental plane.}
\end{table}
%\noindent A Kolmogorov-Smirnov test was also performed on a normalized version of the data sets\footnote{The two-sample KS test is used to test whether two probability distributions differ.}. The null hypothesis (that the data sets come from the same distribution) is rejected at level
%\begin{equation}
%KS>c(\alpha) \sqrt{\frac{n_1+n_2}{n_1 n_2}}
%\end{equation}

%\noindent where $\alpha$ is the confidence level and $n_1$ and $n_2$ correspond to the size of the compared data sets. For a 5\% confidence level test, c(.05)=1.36 and $n_1=n_2=45$ which leads to a KS-statistic threshold of 0.29. From the KS test, the null hypothesis that the data sets come from the same distribution is rejected for the $log E_peak$-Distance distribution.
%\newpage
In order to visually inspect these relations further, we plot 3D histograms of the parameters and distance from the plane in Fig. 9.
These plots strengthen the conclusion that there is no underlying relationship between the studied variables. This outcome reinforces the
previous result that the fundamental plane is independent from the prompt and afterglow emission parameters tested.

\section{Conclusions}\label{discussion}

In our investigation of GRB subclasses, we confirmed the results of Dainotti et al. (2016). Plateau phase GRBs can be used to isolate
a subclass of events that define a very tight plane in a 3D space of $(\log L_{a}, \log T_{a}, \log L_{peak})$. We
confirm that the scatter about this 3D plane is still the smallest when the gold sample, a specific class of GRBs without steep
plateaus and with good coverage of the data, is used for the fitting. The previous gold sample was extended to contain pertinent events
up to 2016 July, for a total of 45 events, and an updated fundamental plane was found with an intrinsic scatter compatible within $1$
$\sigma$ with the previous finding. All other tested relation planes related to the different categories have a larger intrinsic scatter
than the fundamental plane derived from the gold sample. We find that the relation planes for each of the
  mentioned categories are not statistically different from the plane derived from the gold sample, with the exception of the SEE GRBs, which are hence identified as a physically distinct class of objects. Thus, the distance of any particular GRB category from this plane becomes a key parameter, as it can be used as a discriminant feature among long GRBs and the SEE GRBs. 

In addition, we confirmed this 3D relation by using GRBs observed at high energy, namely, by computing the $L_{peak}$ values derived from the {\it Fermi}-GBM, thus showing that the relation is independent of the energy range. The gold fundamental plane obtained with the GBM data presents an intrinsic scatter that is $27\%$ smaller than the gold sample with the {\it Swift} data thus further confirming the existence of this 3D relation. Furthermore, we computed the several category planes by using $T_a$ as a dependent parameter obtaining for each category smaller intrinsic scatters (reaching a reduction of $24\%$ for all the long GRBs).
In order to gain insight into the robustness of the fundamental plane, we explored possible dependencies on a variety of fourth parameters. We found no significant such 4D relations when considering several relevant prompt and afterglow parameters, namely, $\alpha$, $T^{*}_{90}$, $E^{*}_{peak}$, and $\theta_{jet}$, using the $E_{iso}-E_{\gamma}$ relation of the method in Pescalli et al. (2015). 
Given the approximately inverse relation between $T^{*}_a$ and $L_{peak}$ in the afterglow parameters, the plane obtained suggests to first order a strong energy coupling between the afterglow phase (of total energy $E_{aft} \approx T^{*}_a L_{peak}$) and the prompt $L_{peak}$ of the type $E_{aft} \propto L_{peak}^{2/3}$, given the coefficient of $L_{peak}$ in the plane relation of $b=0.64 \pm 0.11$. This energy coupling would favor intrinsic progenitor-based GRB afterglow models over explanations based on external enviromental properties. However, previous studies (Del Vecchio et al. 2016) have shown that there is a dependence of $\alpha$ on the $L_a-T^{*}_a$ relation, which indeed disappears when we consider 4D correlation.
In consistency with the above, we found no significant relation of $\alpha$ as a fourth parameter.
The careful inference of GRB opening angles, as well as their relevance to the 3D correlation between light curve physical parameters explored here, is clearly a desirable extension.

\section{Acknowledgments}
This work made use of data supplied by the UK {\it Swift} Science Data Centre at the University of Leicester. We are particularly grateful
to M. Ostrowski and L. Amati for their relevant comments and suggestions, which helped improving the manuscript. We are thankful to C. Gilberson for his initial contribution to the manuscript during his summer internship at SLAC in 2016. S. S. acknowledges the U.S. Government Department of Energy (DoE), Office of Science, under Contract DE-AC02-76SF00515 and the DOE SULI Internship Program (SULI program) and the Department of Physics of Stanford University for hosting her. We are very grateful to E. Cuellar for coordinating the efforts of the SULI students in this project. M.G.D. acknowledges the Marie Curie Program, because the research leading to these results has received funding from the European Union Seventh Framework Program (FP7-2007/2013) under grant agreement no. 626267. X.N. acknowledges support from UNAM DGAPA PAPIIT grant IN-104517. S.N. acknowledges the funding support of the JSPS, the Mitsubishi Foundation, Associate Chief Scientist Program of RIKEN, a RIKEN pioneering project ``Interdisciplinary Theoretical Science (iTHES)'', and 'Interdisciplinary Theoretical \& Mathematical Science Program (iTHEMS)' of RIKEN.

\appendix
\section{Appendix 1}
For simplicity the luminosity and energy output of GRBs are calculated assuming isotropic emission because the necessary quantity to account for the collimation of the emission, namely, the jet opening angle, is very difficult to acquire without simultaneous multiwavelength observations. While several methods have been used to obtain this angle (Ghirlanda et al. 2004; Goldstein et al. (2016); Lu et al. 2012; Fong et al. 2015), we do not have this estimate for all GRBs. Thus, in order to obtain estimates for $\theta_{jet}$, we turn to the method of Pescalli et al. (2015) in which the jet opening angle can be derived using the $E_{peak}-E_\gamma$ relation (Ghirlanda et al. 2004) and the $E_{peak}-E_{iso}$ (Amati et al. 2002,2009) relation. We use equation (8) of Pescalli et al. (2015), repeated below, to compute these angle values:
\begin{equation}
1-cos \theta_{jet}=(\frac{k_A}{k_G})^{1/G} E_{iso}^{\frac{A-G}{G}}
\end{equation}

\noindent
where $k_A$ and $k_G$ are the normalization constants and $A$ and $G$ are the slopes of the Amati and Ghirlanda relations, respectively. %These parameters are taken fromcan be found from the Amati relation (Amati et al. 2014) and $k_g$, and $G$ can be found from the Ghirlanda relation (Ghirlanda et al. 2004)

\noindent
From Ghirlanda et al. (2004),
\begin{equation}
E_{peak}=k_G \times E_\gamma^G=267 \times (E_{iso} (1-cos \theta_{jet})/(4.3 \times 10^{50} erg))^{0.706 \pm 0.047}
\end{equation}

\noindent
From Amati (2014),
\begin{equation}
log E_{peak}=0.52 \pm 0.06 \times log (E_{iso}/ 10^{52} erg) + 2
\end{equation}
\noindent
or
\begin{equation}
E_{peak}=k_A \times E_{iso}^A=100 \times (E_{iso}/ 10^{52} erg)^{0.52 \pm .06}
\end{equation}

\noindent
By equating and simplifying these equations, one gets the following:
\begin{equation}
1-cos \theta_{jet}\approx\frac{5.36 \times10^{11}}{E_{iso}^{0.26}}
\end{equation}

Thus, an angle can be easily calculated for all GRBs with known $E_{iso}$ values. These values of $\theta_{jet}$ are computed in the current paper. However, we here stress that this estimation is rough. In fact, to effectively and properly use the Amati and Ghirlanda relations to estimate the jet opening angle, the distribution of the scatter must be taken into account, as well as the correlation between the two relations themselves since they both rely on the same $E_{peak}$ values. However, here we use this first-order approach in a first attempt to investigate whether a dependence on the angle is present.

%Figure out how to do appendix a and b

\section{Appendix 2}
When creating the gold sample, we required that the angle of the plateau be less than 41$\degree$ because the distribution of the angles shown in Fig. 10 presents a tail (marked in violet color) above 41\degree. Indeed, if we remove this tail we are able to fit the distribution with a gaussian centered around 25\degree. We also note that this cut excludes only 9\% of the total distribution. This result is compatible with the cut on the previous sample which excluded the 11\% of the data sample.

\begin{figure}[H]
\centering
\includegraphics[width=0.5\hsize,height=0.38\textwidth,angle=0,clip]{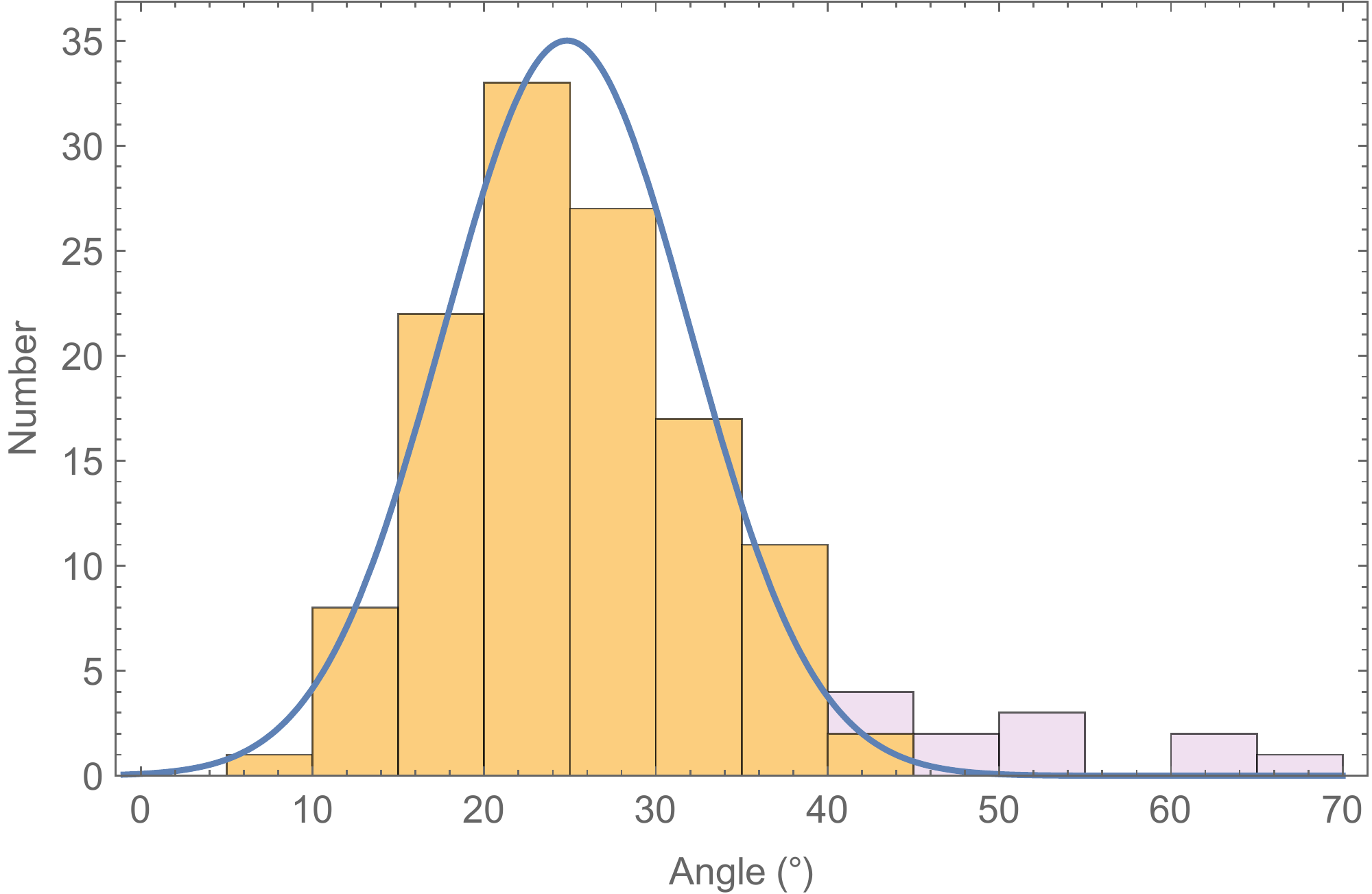}
\caption{Distribution of the plateau angles. Orange represents the GRBs whose plateau angle is within the 41$\degree$ requirement.}
\end{figure}

\section{Appendix 3}
The smoothed histograms are based on a smooth kernel density estimate computed with Mathematica 11.1 with a default build-in option that allows for an automatic bandwith. More specifically, the probability density function for SmoothKernelDistribution for a value {\it x} is given by a linearly interpolated version of for a smoothing kernel {\it k(x)} and bandwidth parameter {\it h} of the form : 
\begin{equation}
1/nh \sum_{i=1}^{n}k(\frac{x-x_i}{h})
\end{equation}
 By default the Gaussian kernel is used.

%\newpage
%\appendix

\end{document}